\documentclass[twocolumn]{aastex631}

\newcommand{\vsys}{$v_{\rm{sys}}$}
\newcommand{\kp}{$K_{\rm{p}}$}
\usepackage{hyperref}

\begin{document}

\title{First Results from WINERED: Detection of Emission Lines from \\ Neutral Iron and a Combined Set of Trace Species on the Dayside of WASP-189 b}
\shorttitle{HRS with WINERED of WASP-189 b's dayside}

\author[0000-0002-9610-3367]{Lennart van Sluijs}
\affiliation{Department of Astronomy, University of Michigan, 1085 S. University Ave., Ann Arbor, 48109, MI, USA}
\correspondingauthor{Lennart van Sluijs}
\email{lsluijs@umich.edu}

\author[0000-0003-3963-9672]{Emily Rauscher}
\affiliation{Department of Astronomy, University of Michigan, 1085 S. University Ave., Ann Arbor, 48109, MI, USA}

\author[0000-0002-1337-9051]{Eliza M.-R. Kempton}
\affiliation{Department of Astronomy, University of Maryland, College Park, MD, USA}

\author[0000-0002-2984-3250]{Thomas Kennedy}
\affil{Department of Astronomy, University of Michigan, 1085 S. University Ave., Ann Arbor, 48109, MI, USA}

\author[0000-0003-0217-3880]{Isaac Malsky}
\affil{Jet Propulsion Laboratory, California Institute of Technology, Pasadena, CA 91109, USA}

\author{Noriyuki Matsunaga}
\affil{Department of Astronomy, School of Science, The University of Tokyo, 7-3-1 Hongo, Bunkyo-ku, Tokyo 113-0033, Japan}
\affil{Laboratory of Infrared High-resolution spectroscopy (LiH), Koyama Astronomical Observatory, Kyoto Sangyo University, Motoyama, Kamigamo, Kita-ku, Kyoto 603-8555, Japan}

\author{Michael Meyer}
\affil{Department of Astronomy, University of Michigan, 1085 S. University Ave., Ann Arbor, 48109, MI, USA}

\author{Andrew McWilliam}
\affil{The Observatories of the Carnegie Institution for Science, 813 Santa Barbara Street, Pasadena, CA 91101}

\author[0000-0002-3380-3307]{John D. Monnier}
\affil{Department of Astronomy, University of Michigan, 1085 S. University Ave., Ann Arbor, 48109, MI, USA}

\author{Shogo Otsubo}
\affil{Laboratory of Infrared High-resolution spectroscopy (LiH), Koyama Astronomical Observatory, Kyoto Sangyo University, Motoyama, Kamigamo, Kita-ku, Kyoto 603-8555, Japan}

\author{Yuki Sarugaku}
\affil{Laboratory of Infrared High-resolution spectroscopy (LiH), Koyama Astronomical Observatory, Kyoto Sangyo University, Motoyama, Kamigamo, Kita-ku, Kyoto 603-8555, Japan}

\author{Tomomi Takeuchi}
\affil{Laboratory of Infrared High-resolution spectroscopy (LiH), Koyama Astronomical Observatory, Kyoto Sangyo University, Motoyama, Kamigamo, Kita-ku, Kyoto 603-8555, Japan}

\begin{abstract}
Ground and space-based observations have revealed that Ultra Hot Jupiters (UHJs,~$T_{\rm{eq}} > 2200 \ \rm{K}$) typically have inverted thermal profiles, while cooler hot Jupiters have non-inverted ones. This shift is theorized due to the onset of strong optical absorbers like metal oxides (e.g., TiO, VO), metal hydrides (e.g. FeH), atomic species (e.g., Fe, Ti), and ions (e.g., H$^-$). High-resolution spectroscopy is valuable for characterizing the thermal, chemical, and dynamical atmospheric structures due to its sensitivity to detailed spectral line shapes. The newly commissioned WINERED high-resolution spectrograph ($R\sim68,000$) on the Magellan Clay 6.5 m telescope enhances capabilities with its high throughput in the J-band (1.13-1.35 $\mu$m), capturing strong spectral features from key atmospheric species. In this study, we report detecting the dayside atmosphere of the UHJ WASP-189 b at a $S/N\sim10$, marking the first exoplanet atmosphere detection in emission with WINERED. Individually, we identify strong neutral iron (Fe) emission lines at a S/N=6.3, and tentatively detect neutral magnesium (Mg) and silicon (Si) at a $S/N>4$. Although not individually detected, we detect a combined set of trace species at a S/N=7.2, which is attributed mostly to neutral chromium (Cr) and aluminum (Al), alongside magnesium and silicon. These results help refine the understanding of key atmospheric species that influence the thermal structure of WASP-189 b and UHJs more broadly.
\end{abstract}


\section{Introduction} \label{sec:intro}
Ultra Hot Jupiters (UHJs, $T_{\rm{eq}} > 2200$ K) provide an opportunity for detailed characterization of exoplanet atmospheres with current observing facilities \citep[e.g.][]{Fortney2021}. Their presumed synchronized orbits \citep{Rasio1996, Showman2002} result in permanent day and night sides that, together with the intense stellar heating, drive strong winds and jets. High dayside temperatures allow for dissociation of both refractory and volatile species from molecules into their atomic and ionic counterparts. Thus, they provide an observational test bed for extreme atmospheric physics not found within the Solar System.
Hot gas giants have been successfully characterized in emission and transmission by space-based low-resolution spectroscopy using JWST and HST \citep[e.g.][]{Huitson2012, Sing2013, Deming2013, Spake2021, Feinstein2023, Ahrer2023, Evans2023, Coulombe2023} and ground-based high-resolution spectroscopy (HRS) \citep[e.g.][]{Snellen2010, Brogi2012, Birkby2013, Hoeijmakers2020b, Wardenier2021, Kesseli2022, Bello-Arufe2022, Prinoth2022, Silva2022, Borsato2023, Gandhi2023, Giacobbe2021, vanSluijs2023, Pelletier2023}. An advantage of high spectral resolution is that the individual spectral lines are resolved, and their line shape \citep[e.g.][]{Snellen2014, Brogi2016}, wavelength offset \citep[e.g.][]{Snellen2010, Brogi2016, Wardenier2021, Kesseli2021, Gandhi2022, Pino2022, Brogi2022, Nortmann2024, Simonnin2024}, and line contrast \citep{Herman2022, Hoeijmakers2022, Pino2022, vanSluijs2023, Lesjak2024}, provide a rich amount of information on the planet's chemistry, thermal structure, rotation, dynamics, and magnetism. Moreover, they encode the three-dimensional (3D) structure, that is, the regional variations as a function of height, latitude, and longitude in the atmosphere \citep[e.g.][]{Beltz2022, Beltz2023, Beltz2024, Wardenier2023, Wardenier2025}.
Hot gas giant observations have revealed a transition around $T_{\rm{eq}}\sim1900$~K, where the cooler hot Jupiters (HJs) shift from non-inverted atmospheres, characterized by spectral absorption features, to inverted atmospheres for the UHJs, which exhibit emission features \citep[e.g.][]{Petz2025}. This transition is theorized to occur due to the onset of strong optical absorbers such as metal oxides (e.g., TiO, VO), metal hydrides (e.g., FeH), atomic species (e.g., Fe, Ti), and ions (e.g., H$^-$) \citep[e.g.,][]{Hubeny2003, Fortney2008, Arcangeli2018, Lothringer2018, Parmentier2018}.
Although the observational evidence for this transition is robust, our understanding of how each optical absorber contributes is still developing. Neutral iron has been consistently detected in many exoplanet atmospheres and has been used to constrain their thermal structure \citep[e.g.,][]{Pino2020, Yan2020, Nugroho2020, Gibson2020}. There is also compelling evidence for H$^-$ as a continuum opacity source. This evidence comes indirectly from the detection of hydroxide (OH) \citep[e.g.][]{Nugroho2021, Landman2021} and from reduced water abundances \citep[e.g.][]{Mansfield2020, Gandhi2024}, which are both attributed to water dissociation into OH and H$^-$. Metal oxides, such as TiO and VO, have been conclusively detected in some hot gas giants, but results for the same planet have frequently been inconsistent \citep{Hoeijmakers2015, Evans2016, Sheppard2017, Hoeijmakers2018, Merritt2021, Prinoth2022, Cont2022, Scandariato2023, Borsato2023, Ouyang2023, Guo2024, Hoeijmakers2024, Prinoth2025}. Cold-trapping has been proposed to explain metal oxide absence or depletion; in this scenario, titanium and vanadium-bearing species condense and sink into the deeper atmosphere on the nightside, effectively removing them from the upper parts of the atmosphere that are probed by HRS~\citep{Parmentier2013}. Iron hydride (FeH) is often predicted to be observable assuming equilibrium chemistry, but despite tentative observational evidence \citep{Evans2016}, searches have not led to any conclusive detections \citep{Kesseli2020}.
The recently commissioned WINERED instrument, which has been installed at the 6.5 meter in diameter Clay/Magellan II Telescope 
located at Las Campanas Observatory in Chile \citep{Kondo2015, Ikeda2022, Otsubo2024}, provides a unique opportunity as its J-band (1.13-1.35 $\mu$m) notably covers strong spectral features of metal oxides, iron hydride, iron, among other atomic and ionic species.
So far, only one published study used WINERED to constrain exoplanet atmospheres \citep{Vissapragada2024}, using the Y-band (0.96-1.11 $\mu$m). This band covers the spectral line from the metastable state of helium at 10830 Å, which has been linked to atmospheric escape \citep{Oklopvcic2018}. Their study with WINERED did not detect helium absorption in the transmission spectrum of the ultra-hot Neptune LTT~9779b, but instead provided an upper limit on its mass loss rate.
In addition to its suitable wavelength coverage, this spectrograph has an especially high throughput (up~to~about~$40\%$) given its high spectral resolution ($R=\lambda /\Delta \lambda_{\rm{FWHM}} = 68,000$) compared to established high-resolution spectrographs that have successfully detected and characterized exoplanet atmospheres such as, e.g.: NIRSPEC \citep{McLean1998}, IRD \citep{Kotani2018},  IGRINS \citep{Park2014, Levine2018}, GIANO \citep{Brogi2018}, CARMENES \citep{Quirrenbach2014}, iSHELL \citep{Rayner2016}, SPIRou \citep{Thibault2012}, CRIRES$^+$, \citep{Lavail2021} and ARIES \citep{McCarthy1998}. Generally, the higher the spectral resolution, the more sensitive the data are to 3D exoplanet atmosphere effects, but this often comes at a cost of instantaneous wavelength coverage and instrumental throughput.
\begin{deluxetable*}{l c c r}
\tablecaption{An overview of relevant stellar and planetary parameters, demonstrating WASP-189 b is a typical ultra hot Jupiter.\label{table:w189b_system_params}}
\tablewidth{0pt}
\tablehead{
\colhead{\textbf{Quantity}} & \colhead{\textbf{Symbol}} & \colhead{\textbf{Value}} & \colhead{\textbf{Reference}}
}
\startdata
\quad Stellar radius & $R_{\star}$ & $2.36\pm0.03\,R_{\odot}$ & \citet{Lendl2020} \\ 
\quad Stellar effective temperature & $T_{\star}$ & $8000 \pm 80 \ \text{K}$ & \citet{Lendl2020} \\ 
\hline
\quad Primary transit time & $T_{0}$ & $2458926.5416960^{+0.0000065}_{-0.0000064}$ (BJD) & \citet{Lendl2020} \\ 
\quad Eccentricity & $e$ & 0\tablenotemark{a} & \citet{Anderson2018} \\ 
\quad Inclination & $i$ & $84.03 \pm 0.14 \ \rm{deg}$ & \citet{Lendl2020} \\ 
\quad Orbital period & $P$ & $2.7240338\pm0.0000067\,\rm{d}$ & \citet{Anderson2018} \\ 
\quad Planetary radius & $R_{\text{p}}$ & $1.619 \pm 0.021 \ \rm{R}_{\rm{Jup}}$ & \citet{Lendl2020} \\ 
\quad Gravitational acceleration & $g$ & $18.8^{+2.1}_{-1.8}$ m/s & \citet{Lendl2020} \\
\quad Dayside equilibrium temperature & $T_{\rm{p}}$ & $3353^{+27}_{-34}\,\rm{K}$ & \citet{Lendl2020} \\ 
\hline
\quad System velocity & $V_{\text{sys}}$ & $20.82\pm0.07$ & \citet{Yan2020} \\
\quad Keplerian velocity & $K_{\text{p}}$ & $197^{+15}_{-16}\,\rm{km/s}$\tablenotemark{b} & \citet{Anderson2018} \\
\enddata
\tablenotetext{a}{Assumed by \citet{Anderson2018} while fitting for the transit shape.}
\tablenotetext{b}{Computed from the orbital parameters assuming a Keplerian orbit.}
\end{deluxetable*}
Here we present the first results from two observing nights with WINERED in HIRES-J mode targeting WASP-189 b. These observations are part of a larger survey targeting HJs over a range of irradiation temperatures and masses. WASP-189 b is a typical UHJ, first detected by the WASP-South survey using the transit method \citep{Anderson2018}. They found it orbits a fast-rotating (line-of-sight velocity of $\sim$100 km/s) hot F-star ($T_{\star}\sim 8000\,\rm{K}$) on a polar orbit. Its orbital, stellar, and planetary parameters have been further accurately constrained by CHEOPS follow-up observations \citep[][see Table~\ref{table:w189b_system_params}]{Lendl2020, Deline2022}.
A few HRS studies have observed WASP-189~b to characterize its atmosphere. \citet{Yan2020} used HARPS-N in the optical to search for a range of atoms and ions. They were the first to report neutral iron emission lines in the dayside atmosphere of WASP-189~b, indicating an inverted thermal structure. Analysis of transmission spectra taken by HARPS, HARPS-N, ESPRESSO and MAROON-X have revealed TiO, Ti, Ti$^+$, Fe, Fe$^+$, Cr, Mg, V, Mn, Na, Ca, Ca+, V, Mn, Ni, Sr, Sr$^+$, and Ba$^+$ \citep{Stangret2022, Prinoth2022, Prinoth2023, Prinoth2024}. Further emission HRS studies have confirmed neutral iron emission in the optical \citep{Deibert2024} on the dayside and also detected carbon monoxide in the near-infrared \citep{Yan2022, Lesjak2025}.
In addition, NUV transmission spectroscopy observations at lower resolution (10-100 \AA) with CUTE \citep{Sreejith2023} detect Mg II and find evidence for Fe II in WASP-189 b. They also revealed an extended upper atmosphere beyond the Roche lobe at a significantly higher temperature ($\sim15,000 \ \rm{K}$) than predicted from hydrodynamical models. This is consistent with Non-Local Thermal Equilibrium (NLTE) chemistry in the upper atmosphere of WASP-189 b. Modeling of hot Jupiters has shown NLTE effects can alter the chemistry and thermal conditions of their upper atmospheres \citep[e.g.][]{Young2020, Wright2022} with observational signatures mainly in transmission, but also in emission spectra \citep[e.g.][]{Barman2002, Swain2010, Moses2011, Fossati2021, Borsa2022, Young2024}. HJs commonly orbit early-type A and F stars \citep{Casasayas-Barris2019}, where varying XUV flux environments can further enhance atmospheric photo-ionization \citep{Fossati2018}. WASP-189 b orbits an early type F-star, which likely increases photo-ionization affecting the abundance of ionic species in its atmosphere.
This manuscript is structured in the following way. Section~\ref{sec:observations_and_processing} discusses our observational strategy and data processing. Section~\ref{sec:atmospheric_modeling} describes how our atmospheric modeling is used to compute spectral templates. Section~\ref{sec:hrccs} explains how we combined our reduced data and spectral templates via HRS cross-correlation methodology for atmospheric characterization. Section~\ref{sec:results} presents the results of this analysis. Section~\ref{sec:discussion} contains the discussion of these results. Finally, Section~\ref{sec:conclusions} presents the main findings and some ideas for future research. 
\section{Observations and Data Processing}
\label{sec:observations_and_processing}
\subsection{Observations}
\begin{deluxetable*}{l c r}
\tablecaption{An overview of the observing setup and conditions for both nights. \label{table:observing_setup}}
\tablewidth{0pt}
\tablehead{
\colhead{\textbf{Quantity}} & \colhead{\textbf{Night 1}} & \colhead{\textbf{Night 2}} \\
}
\startdata
\quad $UT_{\rm{start}}$ & 2023-06-08 22:50 & 2024-04-20 00:18 \\
\quad $UT_{\rm{end}}$ & 2023-06-09 02:45 & 2024-04-20 09:42 \\
\quad Instrument mode & HIRES-J & HIRES-J \\
\quad Slit width [$\mu$m] & 100 & 100 \\
\quad Phase coverage & 0.42-0.48 & 0.47-0.59 \\
\quad Number of frames & 125 & 268 \\
\quad Exposure time [s] & 50/60\tablenotemark{a} & 60 \\
\quad Nodding pattern & ABBA & AB \\
\quad Median humidity [\%] & 13.7 & 18.2 \\
\quad Median S/N per frame pair & 222 & 186 \\
\quad Median FWHM [pixel] & 4.6 & 5.1 \\
\enddata
\tablenotetext{a}{Started with 60 s exposures, but changed to 50 s to avoid detector saturation.}
\end{deluxetable*}
We observed our target WASP-189 for two nights using the 6.5 m Clay/Magellan II Telescope 
located at Las Campanas Observatory in Chile. An overview of the observing setups for both nights is shown in Table~\ref{table:observing_setup}. The setup for the two nights differs slightly as we optimize it for HRS of exoplanet atmospheres with WINERED, which is discussed in more detail throughout this section. In both cases, the instrumental mode used was HIRES-J in combination with a 100 $\mu$m slit. This covers wavelengths from 1.13-1.35 $\mu$m at a spectral resolving power of $\sim$68,000 \citep{Ikeda2022}.
\begin{figure*}
    \includegraphics[width=0.51\textwidth]{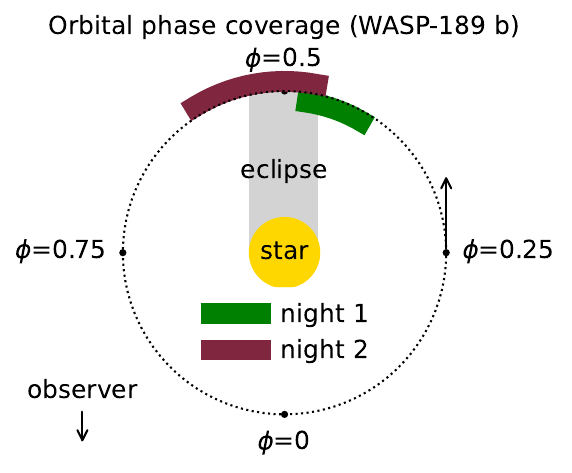}
    \includegraphics[width=0.49\textwidth]{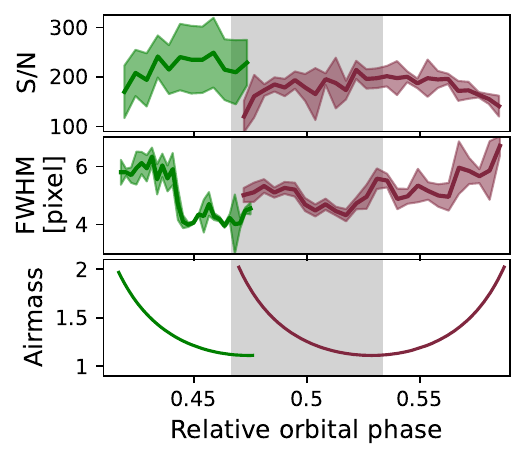}
    \caption{\textit{Left panel:} Phase coverage of WASP-189 b's orbit during our two observing nights. The orbit of WASP-189 b is shown face-on (dotted black line), with the observer's direction indicated. Relative orbital phases ($\phi$) are annotated along quarterly intervals. The first (green) and second (wine red) nights cover the pre-eclipse and post-eclipse phases. Both nights cover (part of) the secondary eclipse. \textit{Right panel:} Time series of three observables: the signal-to-noise rate per frame pair (S/N), Full Width Half Maximum of the trace on the detector (FWHM in pixel), and airmass are shown. The same colors as in the left panel are used to indicate each night. The time series has been binned (ten frames per bin), and the standard deviation of each bin is plotted as the shaded region.} 
    \label{fig:orbital_phase_range} 
\end{figure*}
Combining both nights, our observations cover orbital phases from pre- to post-eclipse (see left panel of Fig.~\ref{fig:orbital_phase_range}). We also observed during the secondary eclipse, since WINERED observing time is allocated nightly. Even though the atmospheric signal of WASP-189~b is obscured by the star during the secondary eclipse, these observations are still useful. They help confirm that the exoplanet signal has disappeared, which aids in distinguishing between actual atmospheric signals and stellar signals or systematic errors.
For each observing night, calibration images were obtained in the morning using the same instrument mode and slit width. This was done following the standard calibration procedure described by \citet{Hamano2024}. First, flat fields are taken from the flat screen without illumination, and then the same images are taken with lighting. Then, flat fields are taken, imaging multiple pinholes. Lastly, a ThAr lamp was used for wavelength calibration of the slits.
The exposure time used differed between the first and second night. During the first night, we started with an exposure time of 60 s, but decided to change it to 50 s to avoid detector saturation towards the end of the observations. This was due to the excellent seeing conditions during that night, and affects 27 frames in total. However, only 4 frames were before the secondary eclipse, mitigating its impact on exoplanet atmospheric detection. During the second night, we used a 50 s exposure time from the start instead. All of the exposure times were in the photon-limited regime, as computed using WINERED's Exposure Time Calculator\footnote{\url{https://merlot.kyoto-su.ac.jp/WINERED/ETC/calc_obstime.php}}. Occasionally, the WINERED instrument tracking software disconnected. In those cases, we had to manually re-center and track the target, after which, observing continued.
Alongside the exposure time, a different nodding pattern was used for each night. The first night used an ABBA nodding pattern. This is a standard nodding pattern used for most HRS exoplanet observations, as it reduces the overhead due to the few seconds it takes in between exposures to nod from slit position~A~to~B. This works well when subsequent exposures at the same slit position are independent. However, this is not necessarily warranted in the case of WINERED, as the NIR detector suffers from persistence. This effectively creates a ghost image of each previous exposure that decays over time. Especially, the persistence of a bright target can be significant in the following exposures of a fainter target. Therefore, exposing twice at the same nodding position can result in an asymmetric persistence signal. Although the persistence of the target itself is expected to be small for many applications, we decided to mitigate its potential risk, considering the high S/N we would like to achieve. This was done by using an AB nodding pattern during the second night, which has a regular time interval between exposures taken at the same slit position. We come back to this point in Section~\ref{sec:observed_instrumental_perfomance}.
\subsection{Data processing}
\label{sec:data_processing}
The raw detector images for both nights were pre-processed using WINERED's processing pipeline {\sc WARP}, which has been made publicly available\footnote{\url{https://github.com/SatoshiHamano/WARP}} by the instrument team and is described in more detail in \citet{Hamano2024}. We used the default input parameters set by the pipeline. The HIRES-J mode covers 22 spectral orders, namely $m=131-153$. We extracted raw spectra over 1.3 times the free spectral range for each spectral order. The pipeline does a bias and dark correction, flat-field correction, and sky subtraction. It then measures the FWHM of the traces on the detector and extracts the raw spectra from the echelle apertures. Alongside the extraction of the raw spectra, the pipeline also produces normalized spectra where a continuum has been fitted and the raw spectra are divided by the fit. The differences between the two spectra are used to compute the continuum S/N per frame pair. The pipeline also includes a wavelength alignment and calibration step, where cross-correlation is used to align and calibrate spectra in the spectral time series. 
\begin{figure}
    \centering
    \includegraphics[width=0.49\textwidth]{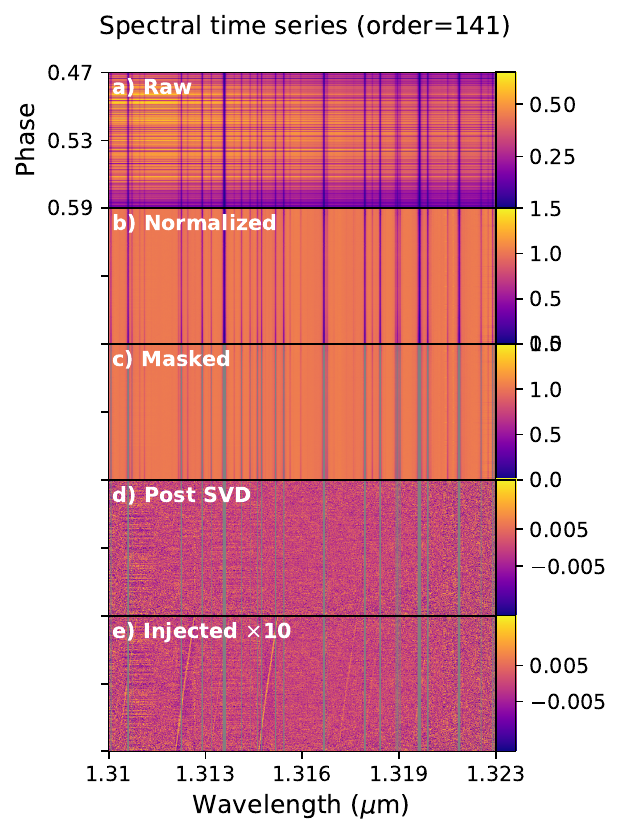}
    \caption{Example of processing a spectral time series for the 141st spectral order and the second night. The color bar indicates photon counts, normalized to an appropriate value for each panel: (\textit{a}) the raw spectral time series is normalized to its maximum value. (\textit{b}) the spectral time series normalized to the continuum. (\textit{c}) the same as in panel 2, but with the strongest telluric features masked, and shown in gray. (\textit{d}) Post removal of the quasi-stationary components by SVD. (\textit{e}) the same as in panel 4, but with the fiducial model injected at ten times the nominal star-to-planet flux ratio. The strongest emission lines are visible as bright sloped lines, blue-shifting due to the planet's orbital acceleration.} 
    \label{fig:processing_steps} 
\end{figure}
The data was further post-processed using our pipeline, which we make available on GitHub\footnote{\url{https://github.com/lennartvansluijs/WINERED}}. This pipeline performs the following steps (see Fig.~\ref{fig:processing_steps}):
\begin{enumerate}
    \item Spectral time series extraction: A multidimensional data cube is created from the WARP output directory. This is done by combining all normalized, wavelength-calibrated spectra of all spectral orders. This data cube thus has the following dimensions: \textit{number of frames} $\times$ \textit{number of spectral channels} $\times$ \textit{number of spectral orders}.
    \item Orbital phase calculation: All times are extracted from the FITS headers and converted from Julian Date (JD) to barycentric Julian Date (BJD). We use the ephemeris and period of WASP-189~b listed in Table~\ref{table:w189b_system_params} to compute the relative orbital phase, assuming a circular orbit \citep[following assumptions by][]{Anderson2018, Lendl2020}.
    \item Artificial injection: following \citet{vanSluijs2023}, an artificial exoplanetary signal can optionally be injected at this stage. This is useful to test the sensitivity to different atmospheric models of WASP-189 b for these observations. This is done by multiplication of the normalized spectral time series by $a\times(1+F_{\rm{scaled}})$, where $F_{\rm{scaled}}$ is a scaled spectral template Doppler-shifted to the exoplanet's expected radial velocity (discussed in Section~\ref{sec:hrccs}), and $a$ an additional scale parameter (introduced by \citet{Brogi2019}).
    \item Telluric masking: strong telluric lines are commonly masked in HRS post-processing, as they are difficult to sufficiently remove from the spectra \citep[e.g.][]{Spring2022, vanSluijs2023}. We mask spectral channels with a continuum flux below a threshold of 70$\%$. This threshold is a tradeoff between excluding the spectral regions contaminated by the strongest telluric lines, while retaining broad wavelength coverage essential for robust signal extraction.
    \item Removal of the quasi-stationary components: we use Singular Value Decomposition (SVD) of the spectral time series for each spectral order to remove the quasi-stationary telluric, stellar, and systematic components. The detection S/N varies with the number of removed components; however, section~\ref{sec:svd} and Fig.~\ref{fig:svd_test} of the appendix show that our key results remain robust despite slight changes in this number. Therefore, we fixed the number of removed components to seven in the remainder of this work. 
\end{enumerate}
The residual spectral time series are the continuum-removed exoplanetary spectra buried in the photon-shot noise. These spectra will be used in the subsequent cross-correlation analysis.
\subsection{Observed instrumental performance}
\label{sec:observed_instrumental_perfomance}
The time series of a selection of relevant observables is shown on the right side of Fig.~\ref{fig:orbital_phase_range}. These are the S/N per frame pair, FWHM of the echelle traces, and airmass are shown as a function of relative orbital phase. As expected, the S/N is broadly inversely correlated to the airmass. The standard deviation of the S/N is larger during the first night compared to the second night. A jump in the FWHM is visible during the first observing night, which corresponds to a change in the observing seeing conditions. During the first night, the median S/N per frame is 222, slightly higher than during the second night, which is 186. During the first night, the median humidity was $13.7\%$, within the normal range for Las Campanas Observatory. Unfortunately, the median humidity was not recorded during the first night in the file headers, due to a broken link between the instrument and the telescope systems. However, we did recover the humidity from the weather archive of Las Campanas Observatory\footnote{\url{https://weather.lco.cl/clima/weather/Magellan/weathercalendar.html}}. The median humidity was slightly higher at $18.2\%$. This is supported by the depth of the normalized telluric lines, which are visually deeper during the second night compared to the first night (see Fig.~\ref{fig:normalised_spectral_orders} of the appendix).
\begin{figure}
    \centering
    \includegraphics[width=0.49\textwidth]{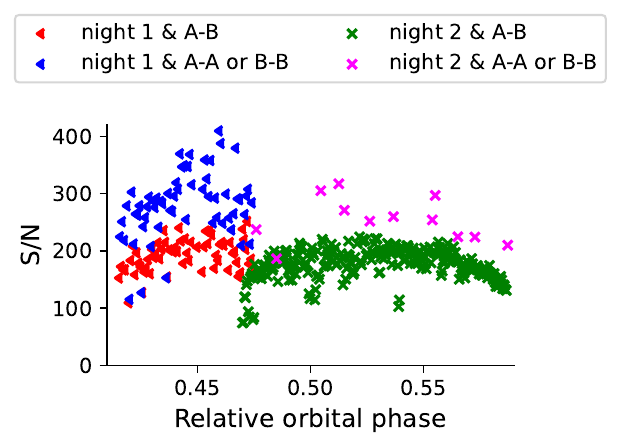}
    \caption{Signal-to-noise ratio (S/N) per frame pair. Triangles pointing to the left indicate frame pairs observed during the first night, and crosses mark the second night. Blue and magenta colors indicate homogeneous pairs, whereas red and green colors indicates heterogeneous combinations. A clear pattern emerges, where the average S/N is higher for homogeneous pairs. We prefer a more stable S/N because our post-processing algorithm can more easily remove quasi-stationary components.} 
    \label{fig:snr_by_nodding_pattern} 
\end{figure}
We investigate whether the difference in standard deviation of the S/N between the first and second nights is due to the different nodding patterns of ABBA and AB, respectively. To explore this, we created two groups: one with A-B pairs and one containing both A-A and B-B pairs. These two groups show two distinct trends in the S/N time series, as seen in Fig.~\ref{fig:snr_by_nodding_pattern}. Due to interruptions of some exposures, a few consecutive frame pairs were taken at the same slit position during the second night, despite having adopted an AB nodding pattern.
Generally pairs taken at the same slit position correlate with a higher S/N value. We speculate three reasons this may be the case. First, it may indicate signal accumulating due to detector persistence when consecutive exposures that are taken at the same slit position. However, it is unlikely persistence alone can explain the magnitude of these effects; the magnitude of the change in S/N is far greater than the effect of persistence as measured from the raw detector images, which is on the order of a few percent. Second, it may be due to systematic errors related to the flat field correction between different nodding positions. Third, it may be due to small pointing offsets after nodding between slit position A and B. Generally, for HRS, a stable S/N is favored, as this improves the removal of quasi-stationary components during SVD. Therefore, we recommend future HRS exoplanet observations with WINERED using AB nodding over ABBA, albeit at a slight cost of extra overhead time to nod between slit positions.
\section{Atmospheric Modeling}
\label{sec:atmospheric_modeling}
Spectral templates need to be computed using atmospheric modeling to compare against the observed spectra, utilizing cross-correlation. We used the {\sc petitRADTRANS} radiative transfer code to compute spectral templates. This code performs radiative transfer by splitting the atmosphere into a stack of atmospheric layers. Our atmospheric model consists of 100 logarithmically equally spaced pressure layers between 10$^{-8}$ and 10$^2$ bar. Each layer has a specified pressure, temperature, mean molecular weight (MMW), and chemical abundances. We adopt the pressure-temperature (P-T) profile observationally constrained by \citet{Yan2020}.
We compute two sets of spectral templates: (1) assuming \textit{chemical equilibrium} and (2) assuming \textit{artificially high chemical abundances}. In the chemical equilibrium case, the MMW and chemical abundances are computed using the {\sc FastChem} chemical network code \citep{Stock2018}. This code assumes solar elemental abundances from \citep{Asplund2009}, and equilibrium chemistry to compute chemical abundances as a function of pressure. 
\begin{figure*}
    \centering    \includegraphics[width=0.9\textwidth]{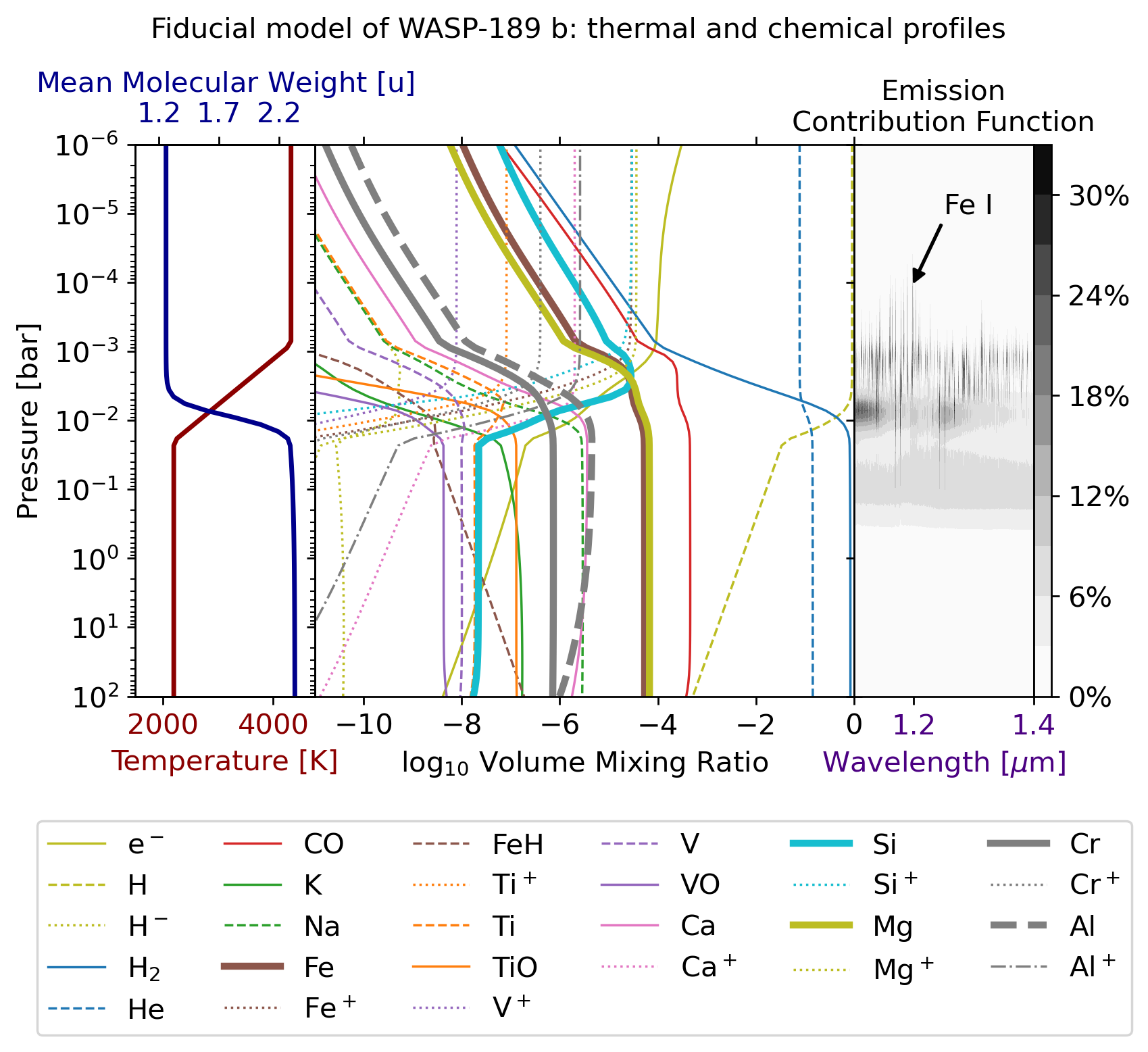}
    \caption{Temperature, MMW, and VMR as a function of pressure for the chemical equilibrium model of WASP-189 b. The P-T profile comes from \citet{Yan2020}. The MMW and VMR values have been computed from the P-T profile using the {\sc FastChem} chemical equilibrium solver, assuming solar elemental abundances. For visual purposes we plot a selection of those most relevant to this work, although chemical abundances of additional atmospheric species were computed with {\sc FastChem} and included in the MMW. \textit{Left panel:} P-T and MMW profiles. \textit{Middle panel:} VMRs for the species shown in the legend on the right. The chemical profiles most relevant to this work (Fe, Si, Cr, Al) have been highlighted using a thicker line width. \textit{Right panel:} the Emission Contribution Function for the fiducial spectral template. A spectral line core of neutral iron is highlighted, which forms at pressures of $\sim0.1-1 \ \rm{mbar}$.}
    \label{fig:fiducial_pt_chemistry} 
\end{figure*}
An overview of the assumed P-T profile, MMW, and chemical abundances is shown in Fig.~\ref{fig:fiducial_pt_chemistry}. We use a constant gravity of $g = 18.8 \ \rm{m/s^{-2}}$ \citep{Lendl2020}. In line with observational constraints of the host star, we assume a $1\times$ solar metallicity atmosphere \citep{Anderson2018}. In the high chemical abundance case, the chemical abundances are artificially raised to $\log VMR=-4$. This is particularly useful to search for opacity sources where equilibrium chemistry predicts low chemical abundance, but that still have strong spectral features in the J-band (e.g. H$_2$O, FeH).
We explore spectral templates containing one of the following opacity sources: CO \citep{Rothman2010}, H$_2$O \citep{Polyansky2018}, OH \citep{Yousefi2018}, K, Na, (both from VALD, Allard wings, see \citet{Molliere2019}), FeH \citep{Wende2010}, TiO \citep{McKemmish2019}, VO \citep{Bowesman2024}, Fe, Fe$^+$, V$^+$, V, Ti$^+$, Ti, Ca, Ca$^+$, Si, Mg, Mg$^+$, Cr, N, and Al (all by Kurucz\footnote{\url{http://kurucz.harvard.edu/}}). All include collision-induced absorption from H-He \citep{Borysow1988, Borysow1989a, Borysow1989b} and H-H \citep{Borysow2001, Borysow2002}. The opacity sources were selected primarily based on their spectral features within the J-band and availability in {\sc petitRADTRANS}. Exceptions were VO, Ti$^+$, whose latest line lists were unavailable in the {\sc petitRADTRANS} format. We acquired the latest hyVO line list through private communication (see acknowledgments) and used the {\sc POSEIDON} \citep{MacDonald2017, MacDonald2023} radiative transfer code for Ti$^+$, retaining the same P-T structure and chemical profile. Alongside spectral templates \textit{with a single opacity source}, we also computed a \textit{fiducial} spectral template that includes all opacity sources assuming equilibrium chemistry. Additionally, templates \textit{with one opacity source removed from the fiducial model} were computed.
\begin{figure}
    \centering
    \includegraphics[width=0.49\textwidth]{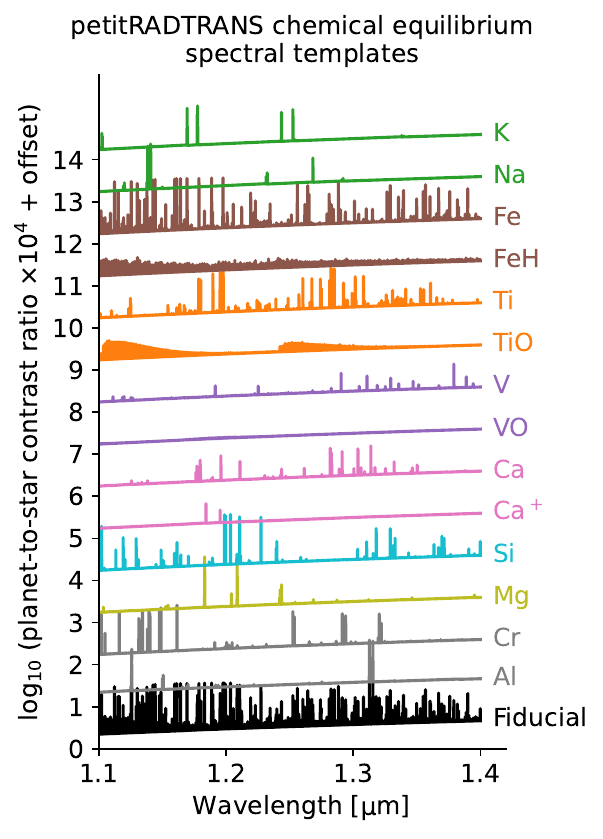}
    \caption{Equilibrium chemistry spectral templates computed using {\sc petitRADTRANS} for the fiducial and single-specie models over WINERED's J-band. These spectral templates do not include H$^-$ as a continuum opacity source. Spectra are offset by unity on the y-scale. Neutral iron has many strong lines and is prevalent in the fiducial model.} \label{fig:prt_spectral_templates} 
\end{figure}
Fig.~\ref{fig:prt_spectral_templates} shows the computed equilibrium chemistry spectral templates for the fiducial model and for those with a single opacity source included. The plots exclude the equilibrium chemistry spectral templates for N, H$_2$O, CO, Ti$^+$, Mg$^+$, Fe$^+$ and V$^+$, since their spectra were featureless, and we excluded these spectral templates from further analysis in this work. The figure shows neutral titanium and iron have a large number of strong lines in the J-band, enhancing their potential detectability. TiO, FeH, and VO have a dense forest of weaker spectral lines. The other atmospheric species range from a few to a moderate number of strong lines. Finally, to understand the pressure range probed, we computed the emission contribution function for the fiducial spectral template with {\sc petitRADTRANS} (see right panel of Fig.~\ref{fig:fiducial_pt_chemistry}). We find typical pressures probed by neutral iron line cores are between $0.1 - 1$ mbar, whereas the continuum forms at deeper pressures around 1 bar.
\section{Cross-Correlation Methodology}
\label{sec:hrccs}
To detect atmospheric species in WASP-189 b we use the signal-to-noise (S/N) method described in detail by \citet{vanSluijs2023}, and which we briefly summarize here. 
\begin{enumerate}
    \item {Model scaling relative to the star: The spectral template is scaled to the expected planet-to-star contrast ratio using
    \begin{equation}
    {F_{\rm{model,scaled}} = \frac{F_{\rm{model}}}{B_{\star}} \left(\frac{R_{\rm{p}}}{R_{\star}}\right)^2},
    \end{equation}
where $F_{\rm{model}}$ is the spectral template, $B_{\star}$ a black body curve at the stellar temperature, $R_{\rm{p}}$ the planet's radius and $R_{\star}$ the stellar radius. The division by a black body at the stellar temperature instead of a stellar atmospheric model is warranted by the hot temperature ($\sim$8000 K), and fast rotation ($\sim$100 km/s) of the host star, suppressing strong stellar spectral features.
}
\item {Spectral resolving power: Each spectral template is convolved to WINERED/HIRES-J's spectral resolution of $R=68,000$.}
\item {Cross-Correlation Function: The residual spectra are cross-correlated against the scaled spectral template for a range of Doppler shifts. We explore a grid of radial velocity offsets between -500 and 500~km/s in steps of 4.4~km/s, set by the spectral resolving power of WINERED/HIRES-J}
\item {Trial rest frames: the Cross Correlation Function (CCF) time series is shifted for a range of trial orbital solutions. We assume a circular orbit, following assumptions by \citet{Anderson2018} and \citet{Lendl2020}. We explore a grid of system velocities between -50 and 50~km/s and Keplerian velocities between 100 and 300~km/s.
For each trial velocity, we shift to the planet's trial rest frame and add the sum of all the CCFs for each spectral order.}
\item {Spectral order weighting: To combine the CCF time series of all spectral orders, we use a model-dependent weighting scheme. This is necessary for two reasons. First, some spectral templates only have spectral features in certain spectral orders.  This is particularly noticeable, for example, in the case of potassium and sodium, which only have a few strong lines in a few of the spectral orders (see Fig.~\ref{fig:prt_spectral_templates}). Second, the telluric, stellar, and systematic residuals differ per spectral order, and they will not be equally sensitive to an exoplanetary signal. Following \citet{vanSluijs2023}, we use injection-recovery to measure the CCF signal in each of the spectral orders. This is done by computing the residual CCF in the planet's rest frame between the observed data and injected data (third step of section~\ref{sec:data_processing}) at an offset radial velocity of $\Delta$\vsys$=\Delta$\kp$=-30$ km/s and a nominal planet-to-star flux ratio, thus $a=1$. Spectral orders are then weighted according to the strength of the peak residual CCF relative to other spectral orders.}
\item{S/N calculation: The $S/N$ ratio is calculated by dividing the combined CCF's standard deviation from the signal in the trial planet's rest frame for each \vsys and \kp \ combination.}
\end{enumerate}
For detected species ($S/N>5$), we estimate $v_{\rm{sys}}$ and $K_{\rm{p}}$ uncertainties from the $SNR_{\rm{peak}}-1$ contour levels. To measure rotational broadening $\delta v_{\rm{broad}}$, we also fit a Gaussian to the summed CCF in the planet's rest frame. For those spectral templates with a sufficiently high $S/N$, the Doppler trail will be visible directly in the CCF time series. In these cases, we will bin the cross-correlation time series into a suitable number of smaller bins and fit a Gaussian profile to each one. The maximum of the Gaussian corresponds with an average velocity offset for those orbital phases within the bin.
\section{Cross-Correlation Results}
\label{sec:results}
\subsection{S/N maps}
\begin{figure}
    \centering
    \includegraphics[width=0.42\textwidth]{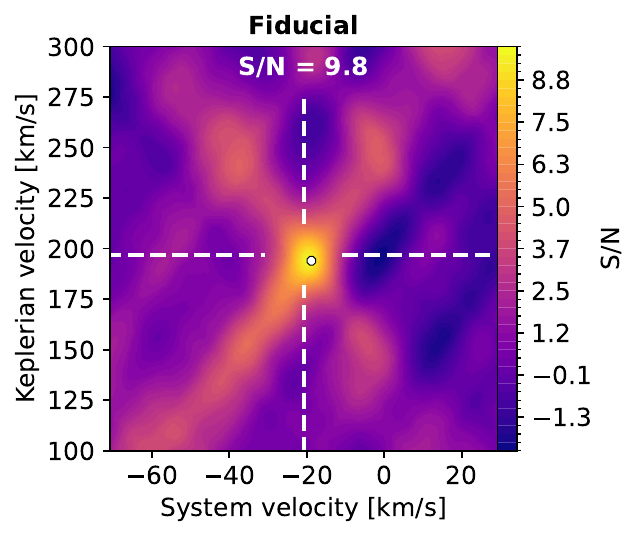}
    \includegraphics[width=0.42\textwidth]{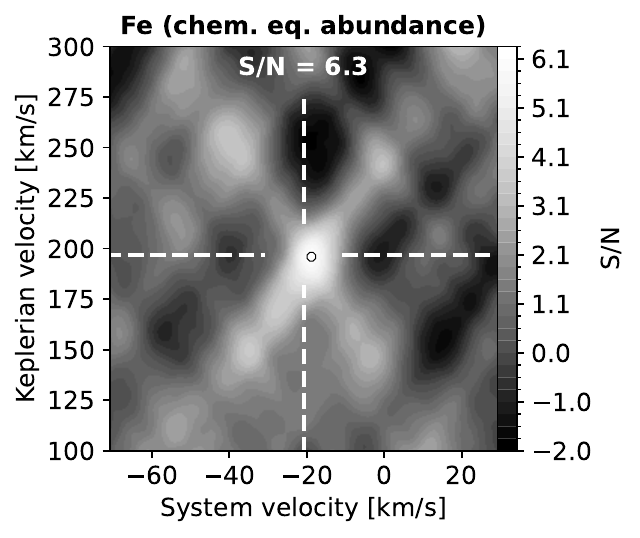}
    \includegraphics[width=0.42\textwidth]{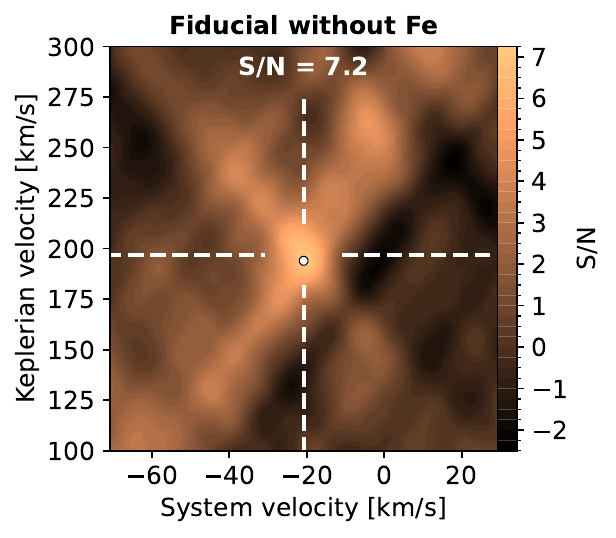}
    \caption{S/N as a function of system velocity ($v_{\rm{sys}}$) and Keplerian velocity ($K_{\rm{p}}$) when combining all out-of-eclipse phases for both nights in the barycentric rest frame. The literature values for $v_{\rm{sys}}$ and $K_{\rm{p}}$ are indicated by the white dashed cross. The maximum S/N is marked by the small circle, and its value is annotated at the top of the plot. \textit{Top left:} for the fiducial spectral template \textit{Top left:} for the neutral iron (Fe) spectral template using the chemical profile assuming equilibrium chemistry. \textit{Bottom:} the fiducial spectral template without neutral iron.}
    \label{fig:snr_detections}
\end{figure}
The S/N maps with detections are shown in the panels of Fig.~\ref{fig:snr_detections}. The same maps are shown for each night separately in Fig.~\ref{fig:snr_detections_each_night} of the appendix. They show a significant detection of the atmosphere of \hbox{WASP-189 b} (fiducial model is detected at $\rm{S/N=9.8}$), and neutral iron ($S/N=6.3$). These detections are well above the limit for which correlated noise in these types of S/N-maps is known to stochastically cause false positives $S/N < 4$ \citep{Cabot2020, Spring2022}. Our peak $S/N$-value is also significantly higher than the absolute minimum value in both maps, which indicates it is unlikely to be caused by correlated noise.
The location in the S/N maps aligns well with the literature values $v_{\rm{sys}}=20.82\pm0.07\ \rm{km/s}$ \citep{Yan2020} and $K_{\rm{p}}=197^{+15}_{-16}\ \rm{km/s}$ \citep{Anderson2018}. For the fiducial model we find $v_{\rm{sys}}=-19.5^{+2.4}_{-2.6} \ \rm{km/s}$ and $K_{\rm{p}}=194^{+7}_{-6} \ \rm{km/s}$, and for the iron-only model we find $v_{\rm{sys}}=-19.0^{+3.1}_{-2.0} \ \rm{km/s}$ and $K_{\rm{p}}=195^{+8}_{-9} \ \rm{km/s}$. For the fiducial model, we find $\delta v_{\rm{broad}} = 6.1 \pm 0.4 \ \rm{km/s}$ and for the iron-only we find $\delta v_{\rm{broad}} = 5.4 \pm 0.2 \ \rm{km/s}$
The \vsys versus \kp \ maps for our other spectral templates are shown in the appendix: Fig.~\ref{fig:snr_other_species} shows the non-detected species assuming chemical equilibrium, and Fig.~\ref{fig:snr_high_chem_abund} shows all species assuming high chemical abundance ($\log VMR = -4$). Except for the high chemical abundance iron-only template, all of these are below our tentative detection limits of $S/N < 4$. The only exception is silicon (Si), which is tentatively detected at a $S/N=4.1$.
\begin{figure}
    \centering
    \includegraphics[width=0.45\textwidth]{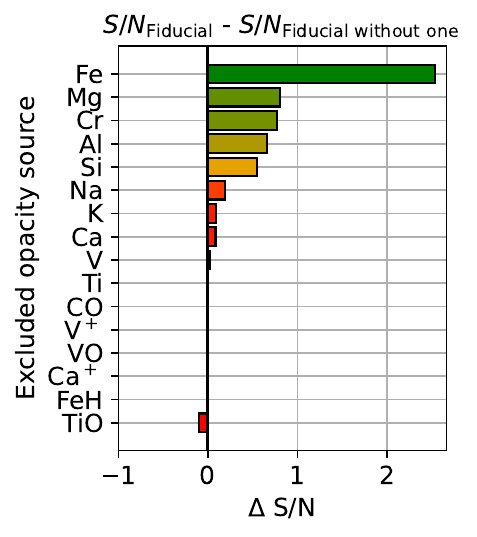}
    \caption{Differences in signal-to-noise ratio ($\Delta S/N$) are measured between the complete fiducial spectral template and templates omitting one opacity source at a time. Large positive $\Delta S/N > 1$ values indicate a significant detection of the excluded opacity source (e.g., Fe), whereas near-zero values suggest that excluding or including the opacity source has little effect, indicating insignificant detection.} 
    \label{fig:snr_diff_fiducial_without_one}
\end{figure}
Despite the non-detections of other species than iron, the fiducial model is detected significantly higher than the iron-only model, suggesting a contribution from a subset of the other atmospheric species. To investigate this further, we also cross-correlated the data against the fiducial spectral template that excluded a single atmospheric species. The difference in $S/N$-ratios is shown in Fig.~\ref{fig:snr_diff_fiducial_without_one}. Excluding iron significantly reduces the detection $S/N$ by about two, indicating its presence in the atmosphere. The exclusion of other species reduces the detection $S/N$ slightly, but by less than one, indicating an insignificant presence or absence. To check if their combined contribution is significant, we also cross-correlated with a fiducial spectral template with iron excluded. This template is significantly detected around WASP-189 b's expected \vsys and \kp \ location at a $S/N=7.2$ (see bottom right of Fig.~\ref{fig:snr_detections}). Thus, the combined contribution of all opacity sources other than iron is significant, despite insignificant detections for each of these species individually. Fig.~\ref{fig:snr_diff_fiducial_without_one} shows that this can mainly be attributed to a combination of magnesium (Mg), chromium (Cr), aluminum (Al), and silicon (Si). 
\subsection{Doppler trails}
\begin{figure*}
    \centering
    \includegraphics[width=0.85\textwidth]{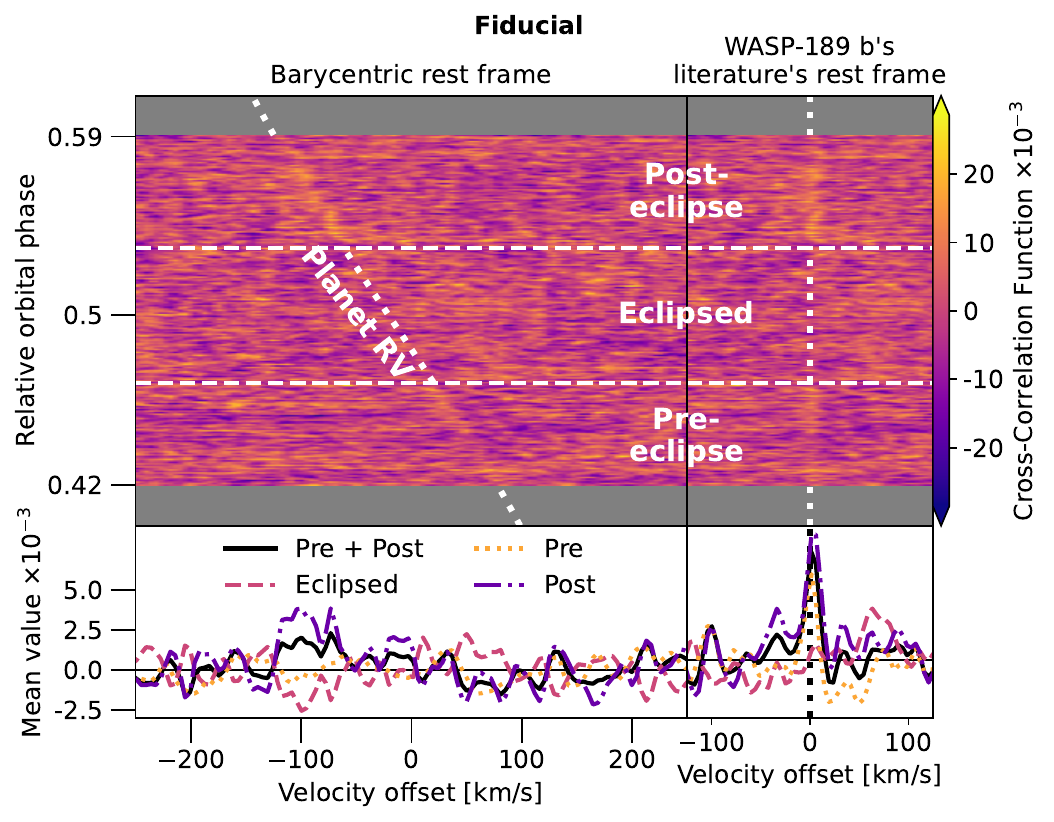}
    \caption{Cross-Correlation Function (CCF) time series for the fiducial spectral template showing the exoplanet's Doppler signature is visible on both nights. \textit{Top rows:} CCF-value in the barycentric rest frame (left) or WASP-189 b's a literature's rest frame (right; assuming the literature values of \kp \ and \vsys \ listed in Table~\ref{table:w189b_system_params}). The planet's expected radial velocity trail (RV) is drawn as a white dotted line. Pre-eclipse, post-eclipse, and eclipsed phases are indicated. \textit{Bottom rows:} Mean values of the CCF in each rest frame for a subset of orbital phases as shown in the legend.} 
    \label{fig:ccf_time_series}
\end{figure*}
The Cross-Correlation Function (CCF) time series is shown for the fiducial model in Fig.~\ref{fig:ccf_time_series}. This figure includes orbital phases observed during the secondary eclipse, which were excluded from the $S/N$ calculation. In the top panels, a peak in the CCF time series, the Doppler trail, can be seen in the pre-eclipse data, disappears during the secondary eclipse, and reappears in the post-eclipse data. This becomes more significant when summing up the CCF during these different orbital phase intervals (bottom panel). This provides unambiguous evidence that the signal originates from WASP-189 b and is not due to telluric, stellar contamination, or systematics. The post-eclipse trail is clearer than the pre-eclipse, and this is reflected by the higher detection S/N during the second night (see Fig.~\ref{fig:snr_detections_each_night} of the appendix). We speculate this may be due to the higher stability of the continuum S/N (see Fig.~\ref{fig:orbital_phase_range}) compared to the first night.
\begin{figure*}
    \centering
    \includegraphics[width=0.95\textwidth]{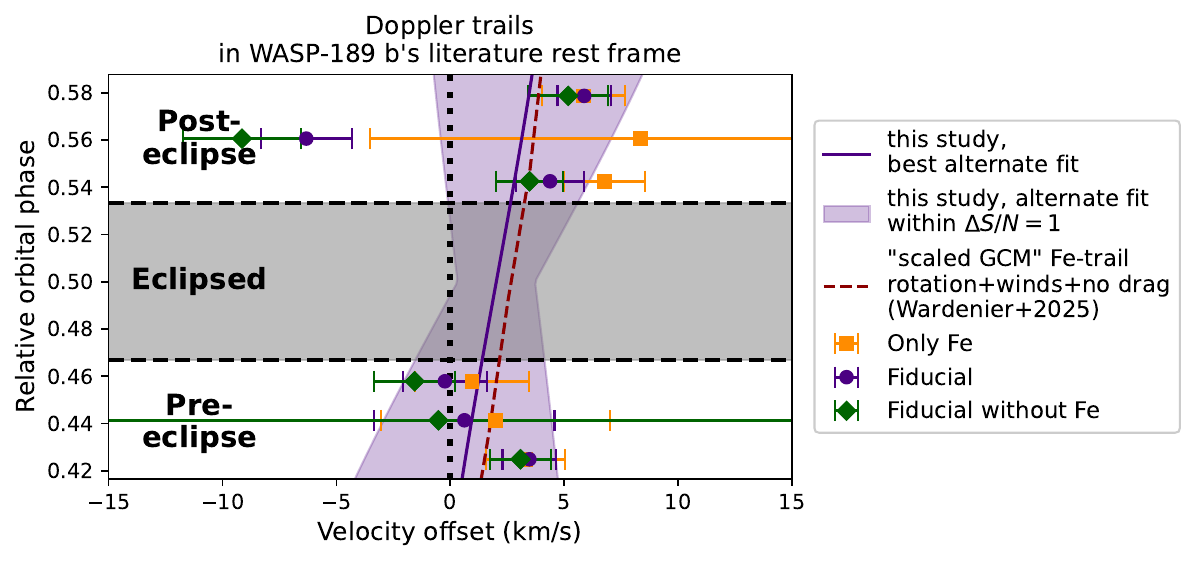}
    \caption{The scatter points mark the velocity offsets resulting from a Gaussian fit to the binned CCF time series in WASP-189 b's literature rest frame (\vsys and \kp~listed in Table~\ref{table:w189b_system_params}). Three bins are used pre-eclipse and post-eclipse. The scatter points' colors correspond to different spectral templates: only-iron (at equilibrium chemical abundance), fiducial without titanium (our highest S/N template) and fiducial without iron and titanium. The purple line indicates our best alternative orbital fit compared to the literature, for the highest S/N spectral template, where the purple shaded region indicates orbital fits consistent within $\Delta S/N = S/N_{\rm{peak}}-S/N_{\rm{velocity \ offset}} = 1$. This is compared against predictions by a GCM (red dashed line) for the velocity offset of iron lines in another typical UHJ, but with the slope scaled to the rotational velocity of WASP-189 b. This GCM model included the effect of rotation, winds, and assumed no drag. Our alternative orbital fit is consistent with the literature, and its additional slope is consistent with 3D effects.} \label{fig:doppler_trails} 
\end{figure*}
We applied the Gaussian fitting to the spectral templates with iron only, the fiducial template, and the fiducial template without iron. Three bins were used pre-eclipse and post-eclipse. Our alternative orbital fit is shown in the literature planet rest frame in Fig.~\ref{fig:doppler_trails}. If our study of WASP-189 b had found the same \vsys and \kp~values as the literature, all points should follow a vertical line at a net-zero velocity offset. However, our \vsys and \kp values differ slightly from the literature value from \citet{Anderson2018}, thus our best alternative fit has a slight slope in Fig.~\ref{fig:doppler_trails}. Although this is consistent with the literature value within our measured error on \vsys and \kp, as indicated by the shaded region, it may also hint towards 3D effects, which we get back to in Section~\ref{sec:3d_effects}.
\section{Discussion}
\label{sec:discussion}
\subsection{Iron detection}
Our detection of neutral iron in the atmosphere of WASP-189 b is the first detection in the J-band. The detection of iron is compatible with equilibrium chemistry, which predicts a relatively high chemical abundance ($\log_{10}X_{\rm{Fe}}\sim-5$) around line-forming pressures ($P\sim10^{-3}-10^{-5} \ \rm{mbar}$). The detection of neutral iron emission lines independently confirms a temperature inversion in the atmosphere of WASP-189 b. This agrees with the observed trend that UHJs have inverted dayside atmospheres \citep[e.g.][]{Petz2024}. Our iron detection is consistent with previous observations from other instruments, including HARPS-N \citep{Yan2020}, CRIRES$^+$ \citep{Lesjak2024}, GHOST \citep{Deibert2024}. Therefore, this work contributes to an emerging trend that iron is detectable in UHJ atmospheres across a wide wavelength range, namely from the optical to the near-infrared. 
\subsection{Titanium chemistry}
\begin{figure*}
    \centering
    \includegraphics[width=0.43\textwidth]{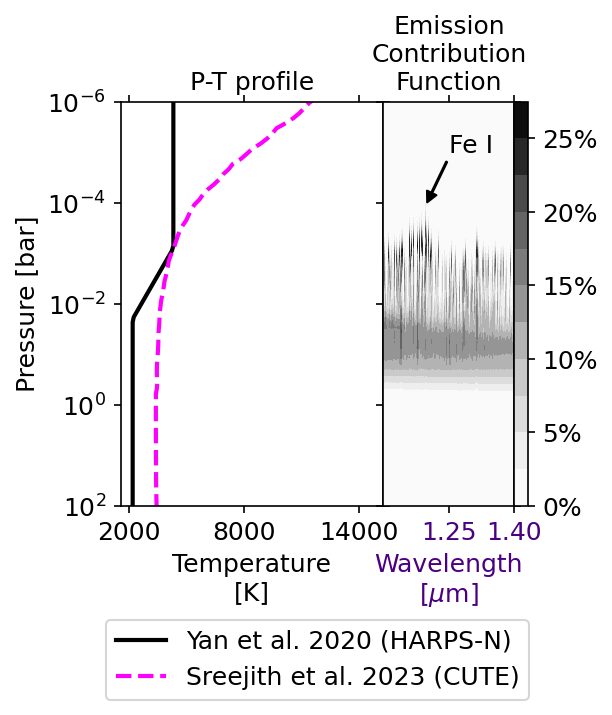}
    \includegraphics[width=0.55\textwidth]{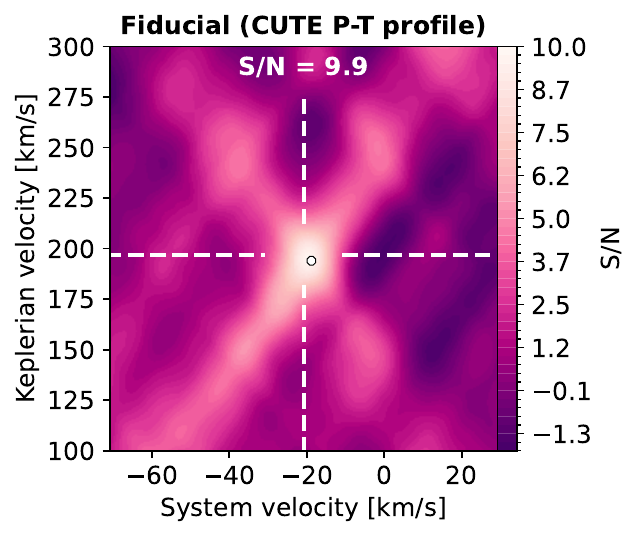}
    \caption{\textit{Left:} the left panel shows a comparison between the P-T profile from \citet{Yan2020} who used HARPS-N emission spectroscopy and the the P-T profile constrained by \citet{Sreejith2023} who used CUTE NUV transit observations. The right panel shows the Emission Contribution Function for the fiducial spectral template using the CUTE P-T profile. A strong neutral iron emission line core is indicated. \textit{Right:} similar to the S/N maps of Fig.~\ref{fig:snr_detections}, but shown for the fiducial spectral template that uses the P-T profile from CUTE and with chemical profiles computed using {\sc FastChem}. This spectral template is detected at a $S/N=9.9$.}
\label{fig:cute}
\end{figure*}

\begin{figure}
    \centering
    \includegraphics[width=0.35\textwidth]{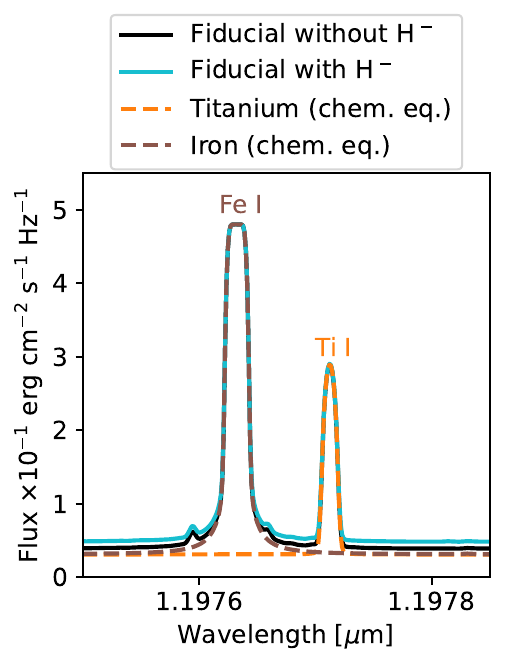}
    \includegraphics[width=0.48\textwidth]{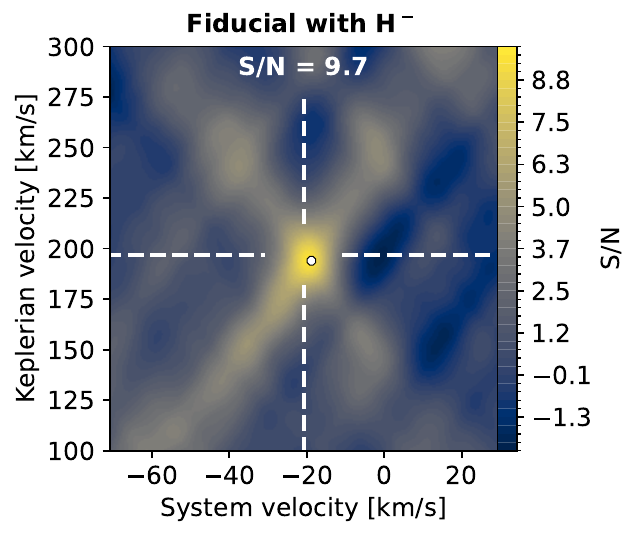}
    \caption{\textit{Top:} zoom of the fiducial spectral template with and without H$^-$ around 1.1977 $\mu$m showing a titanium and iron emission line. The effect on both lines is negligible. \textit{Bottom:} similar to Fig.~\ref{fig:snr_detections}, but for the fiducial spectral template with H$^-$ opacity included, which is detected at a comparable $S/N=9.7$.}
\label{fig:effect_of_H-}
\end{figure}

We notably do not detect Ti$^+$, Ti, or TiO in the emission spectrum of WASP-189 b. These results are consistent with results from GHOST by \citet{Deibert2024} who did not detect titanium-bearing species in the optical emission spectra of WASP-189 b. In contrast, ionized titanium, titanium, and titanium oxide have all been detected in the terminator regions of WASP-189 b using transmission spectroscopy \citep{Prinoth2022, Prinoth2023, Prinoth2024}.
Our injection-recovery tests demonstrate we are not sufficiently sensitive to the titanium oxide and ionized titanium at the chemical abundances from equilibrium chemistry. For TiO, this is attributed to the expected low chemical abundance ($\ logX_{\rm{TiO}}<-7$ at any pressure). For ionized titanium, there are not enough sufficiently strong lines in the J-band to be detectable. In contrast, our injection-recovery test recovers the neutral titanium spectral template at equilibrium chemical abundances at a $S/N = 5.8$. However, this is less than half of the retrieved injection iron-only model $S/N=13.8$. Given that for iron, our observed S/N is about half of the injected S/N, we speculate the observed $S/N$ for titanium to be too low to be detectable.
Despite this, if we assume the non-detection of neutral titanium on the dayside is significant, we discuss four explanations: (1) cold-trapping, (2) increased ionization of titanium, (3) line list inaccuracies, and (4) masking of the titanium lines below the H$^-$ continuum.
In the case of cold-trapping, titanium-bearing species condense on the night side, sinking to deeper parts of the atmosphere that are not probed by emission HRS \citep{Parmentier2013}. This mechanism is disfavored in the case of WASP-189 b for two reasons. First, it would require a circulation pattern with effective vertical mixing around the terminator regions, but without transport to the dayside. Second, for WASP-189 b, the nightside is possibly too hot for cold-trapping; \citet{Deline2022} set a 2250 K upper limit on the nightside temperature, constrained by degeneracies between the geometric albedo and the heat redistribution efficiency. This is above the condensation temperature of neutral titanium \citep{Pelletier2023} at the higher end. 
Another explanation may be that the ionization of titanium is more efficient than expected from equilibrium chemistry, and all the dayside titanium is in an ionized form. To deplete titanium by one order of magnitude, we find that equilibrium chemistry models require an atmosphere that is $\sim$500 K hotter than constrained by \citet{Yan2020} at the pressures where the line cores form. \citet{Yan2020} tightly constrained the upper atmospheric temperature to $T_{\rm{upper}}=4340^{+120}_{-100} \rm{K}$ at pressures below $\log{P_{\rm{upper}}}=-3.10^{+0.23}_{-0.25}$. However, this maximum temperature is set by the conditions in the region where the neutral iron line cores form ($\sim0.1-1$ mbar), where NLTE such as photo-ionization are expected to be negligible. Their observations lack sensitivity at significantly lower pressures, where NUV transmission spectroscopy observations by CUTE have constrained much hotter temperatures \citep[$\sim 15,000$ K, see][]{Sreejith2023}. In any case, we tested this by computing an additional fiducial spectral template using the CUTE constraints on the thermal profile of WASP-189 b (see Fig.~\ref{fig:cute}) and using {\sc FastChem} to compute the chemical profiles. Compared to the P-T profile by \citet{Yan2022}, the CUTE P-T profile is slightly hotter and has a monotonically increasing hotter upper atmosphere. A cross-correlation analysis detects this spectral template at a slightly higher $S/N=9.9$, which is slightly higher, but approximately similar to the fiducial model by \citet{Yan2020} which was detected at a $S/N=9.8$. This comparable $S/N$ may be explained by the similarity of the P-T profiles around the pressures where the iron line cores form ($\sim$0.1 mbar, see Fig.~\ref{fig:cute}). This result indicates that an overall slightly hotter atmosphere, particularly where the continuum forms, is also consistent with the WINERED data; possibly explaining increased titanium depletion. Regardless, ionized titanium was undetected, even using our high chemical abundance spectral template (see Fig.~\ref{fig:snr_high_chem_abund}). Injection-recovery only tentatively detected this spectral template at $S/N=3.9$. This demonstrates that the WINERED data is not sufficiently sensitive to Ti$^+$.
Third, the line lists for the titanium-bearing species may not be sufficiently accurate \citep{Hoeijmakers2015}. We aimed to use the most up-to-date line list of TiO, Ti, and Ti$^+$ available. However, \citet{Prinoth2024} excluded certain wavelength parts of their analysis, due to known inaccuracies in their Ti line list, albeit in the optical regime. \citet{Serindag2021} found that using different TiO line lists can impact previously reported detection significances in the atmosphere of the UHJ WASP-33 b.
Fourth, H$^-$ opacity may effectively obscure emission lines, particularly those forming at deeper pressures than iron lines. Since our original model did not include H$^-$ opacity, we tested its effect by creating a spectral template that incorporated H$^-$ using {\sc petitRADTRANS}, which implements the effects of H$^-$ following~\citet{Gray2008}. Indeed, we find that the H$^-$-continuum mutes all spectral features, including both iron and titanium lines, and the effect is relatively stronger for the lines forming at deeper pressures. However, the effect is negligible relative to our spectral features (see Fig.~\ref{fig:effect_of_H-}), and unsurprisingly, cross-correlation with the fiducial spectral template with H$^-$ indicates a non-significant change in the S/N ratio as the $S/N=9.7$, comparable to the S/N of the fiducial spectral template.
\subsection{Other atmospheric species}
The significant detection ($S/N = 7.2$) of the fiducial spectral template without iron suggests the combined set of Mg, Si, Cr, Al, Na, V, Ca, K, VO, FeH, and TiO is detected. Silicon, magnesium, chromium, and aluminum all show a peak $S/N$ close to the measured \vsys and \kp \ values. These four sources also show the most significant drop in $S/N$ when removed from the fiducial model (see Fig.~\ref{fig:snr_diff_fiducial_without_one}). Alongside this, spectral templates with silicon (both chemical equilibrium and higher chemical abundance) and magnesium (at higher chemical abundance) were tentatively detected, e.g., $S/N>4$. Altogether, these results hint towards tentative signals of silicon, magnesium, chromium, and aluminum in the WINERED data, but are not conclusive from these data alone.
The non-detection of other atmospheric species than neutral iron is consistent with previous findings in the optical and near-infrared. \citet{Yan2020} also searched for Fe$^+$, Ti, Ti$^+$, TiO, and VO, and \citet{Deibert2024} checked for all species included in this work. Both studies resulted in null detections of anything other than iron. However, for Ti$^+$, \citet{Deibert2024} used a binary mask instead of a spectral template, and this null detection may be attributed to this. Similar to our study, the fiducial model from \citet{Deibert2024} was detected at a higher $S/N$-ratio than neutral iron alone ($S/N=7.8$ instead of $S/N=5.5$), indicating the presence of a set of trace species.
\subsection{Hints of 3D effects}
\label{sec:3d_effects}
Our \kp$=194^{+7}_{-6} \ \rm{km/s}$ value is consistent with the literature value of \kp$=197^{+15}_{-16} \ \rm{km/s}$ computed from the radial velocity signature measured by \citet{Anderson2018} assuming a circular orbit. However, our slightly lower \kp-value could also be hinting towards 3D effects. \citet{Zhang2017} predicted that HJ rotation and winds can result in 1-3 km/s net Doppler offsets. \citet{Beltz2022} showed this can result in orbital phase-dependent Doppler shifts of similar magnitude using emission HRS. \citet{Hoeijmakers2024} have pointed out that for a synchronized rotation rate, the measured Keplerian velocity would be lower than the expected velocity due to what parts of the exoplanet's disk are visible. \citet{Wardenier2025} generalize this towards $|\Delta K_{\rm{p}}| < v_{\rm{rot}} + v_{\rm{jet}}$. This results in a slope of our alternate orbital fit when plotted in the literature's planet's rest frame as shown in Fig.~\ref{fig:doppler_trails}. We also over-plot the Doppler shift based on a modified GCM prediction for another UHJ WASP-76~b for iron lines by \citet{Wardenier2025}. This was done by taking the Doppler trail for iron as a function of the orbital phase for the drag-free GCM with Doppler shifts from rotation and winds included, and scaling the slope according to the rotational velocity ratio between WASP-189 b and WASP-76 b (assuming a synchronized orbit). The `scaled GCM' iron Doppler trail is a close match to our alternate orbital fit. We do not find significant evidence for a shift between Doppler trails for the three spectral templates.
Our results from the Doppler shifts are consistent with the Doppler broadening we measured of $\delta v_{\rm{broad}} = 6.2 \pm 0.9 \ \rm{km/s}$ for the fiducial model, $\delta v_{\rm{broad}} = 5.4 \pm 0.2 \ \rm{km/s}$ for the iron-only model, and $\delta v_{\rm{broad}} = 6.9 \pm 0.8 \ \rm{km/s}$ for the fiducial model without iron. This is in line with values derived by \citet{Lesjak2025} who measured day-to-night side winds of 4.4$^{+1.8}_{-1.2}$ km/s and an additional eastward jet of 1.0$^{+0.9}_{-1.8}$ km/s. The measurement of significant rotational broadening, but absence of a strong offset in $\Delta v_{\rm{sys}}$ is consistent with results by \citet{Beltz2021}, who found cross-correlation with a 3D GCM spectral template with and without Doppler effects, mainly increases the broadening in the S/N-map, but did not result in additional the velocity offset.
\section{Conclusions}
\label{sec:conclusions}
In this work, we presented the analysis of two nights of HRS data of the UHJ WASP-189 b, obtained with the recently deployed WINERED spectrograph at the 6.5 m Clay/Magellan II Telescope located at Las Campanas Observatory in Chile. The results establish WINERED as an exoplanet atmosphere characterization instrument in the J-band, bridging the optical and near-infrared regimes.
The main conclusions of this work are:
\begin{enumerate}
\item The atmosphere of WASP-189 b is detected with WINERED/HIRES-J at a high $S/N\sim10$. Previously constrained thermal profiles derived from NIR emission spectroscopy from \citet{Yan2020} ($S/N=9.8$), as well as a slightly hotter thermal profile inferred from NUV transit observations from \citet{Sreejith2023} ($S/N=9.9$), both provide a good match to our data.
\item We detect neutral iron emission lines in the atmosphere of WASP-189 b at a $S/N = 6.3$, indicating an inversion layer. This result is consistent with previous HRS reconnaissance of this planet using different instruments and wavelength bands.
\item We detect the fiducial model without iron at a $S/N=7.2$. This suggests a combined set of trace species, despite not being detectable individually. Comparison with models that excluded one opacity source from our fiducial model hints that this can be mostly attributed to neutral magnesium (Mg), chromium (Cr), aluminum (Al), and silicon (Si). Individually, neutral silicon is tentatively detected in the WINERED data at a $S/N=4.1$, assuming chemical equilibrium, and neutral magnesium at a $S/N=4.3$, assuming an enhanced chemical abundance compared to equilibrium chemistry ($\log VMR=-4$).
\end{enumerate}
Subsidiary conclusions are:
\begin{enumerate}
    \setcounter{enumi}{2}
    \item Using WINERED with AB nodding, we observed a more stable continuum S/N between consecutive frames compared to ABBA nodding. This can likely be attributed to the mitigation of detector persistence, flat field systematics, or pointing errors while guiding on target.
    \item The disappearance of the Doppler trail during the secondary eclipse highlights the benefits of observing (part) of the secondary eclipse, to unambiguously confirm that the signal originates from the planet and cannot be attributed to correlated noise, telluric correction, or the stellar lines.
    \item The measured phase-dependent velocity shifts and velocity broadening are consistent with a 3D circulation where WASP-189 b has a synchronous rotation rate with an additional atmospheric jet.
\end{enumerate}
Finally, we highlight some directions for future research. First, more observations or combining these observations with existing data sets would help to confirm the trace species individually, particularly to verify the tentative detections of silicon and magnesium. Second, this study adds to the emerging evidence that iron can be observed over a wide wavelength range from the optical to the near-infrared. This implies a range of weak and strong spectral lines probing different atmospheric layers and depths. A study combining all of these data will provide high sensitivity to the P-T structure, possibly winds at different atmospheric layers \citep[e.g.][]{Kesseli2024} and trace wavelength-dependent differences in where the spectral continuum forms. Third, multiple nights on an optical spectrograph would be better suited to test the hypothesis of ionized titanium on the dayside of UHJs like WASP-189 b. Finally, our injection-recovery tests show WINERED should be highly sensitive to TiO, FeH, and VO at higher chemical abundances. Targeting hot Jupiters with a cooler dayside, but sufficiently hot nightside to prevent cold-trapping, will increase the likelihood of detecting these molecules with WINERED.
\facilities{LCO(WINERED)}
\software{The software accompanying this work is available on the first author's GitHub\footnote{\url{https://github.com/lennartvansluijs/WINERED-WASP-189-b}} and archived in Zenodo \citep{lennart_van_sluijs_2025_16113076}, Astropy {\citep{Astropy2013, Astropy2018, Astropy2022}},
Matplotlib \citep{Matplotlib2007}, {\sc petitRADTRANS} \citep{Molliere2019}, {\sc {WARP}} \citep{Hamano2024}, {\sc Fastchem} \citep{Stock2018}, Numpy \citep{Harris2020}, Scipy \citep{Virtanen2020}, {\sc POSEIDON} \citep{MacDonald2017, MacDonald2023}}
\begin{acknowledgments}
We would like to thank the anonymous reviewer for their insightful comments which have improved the quality of this work. We would like to thank Sam de Regt for providing the latest hyVO line list in a format suitable for {\sc petitRADTRANS}, and comments via private communication, which improved the quality of this work. This paper is based on WINERED data gathered with the 6.5 m Clay/Magellan II Telescope 
located at Las Campanas Observatory, Chile. WINERED was developed by
the University of Tokyo and the Laboratory of Infrared High-resolution Spectroscopy, Kyoto
Sangyo University, under the financial support of KAKENHI (Nos. 16684001, 20340042, 21840052, and 26287028) and the MEXT Supported Program for the Strategic Research Foundation
at Private Universities (Nos. S0801061 and S1411028). The observing runs in 2023 June
and 2024 April were partly supported by KAKENHI (grant No. 19KK0080) and JSPS Bilateral
Program Number JPJSBP120239909. This work also received financial support from the Heising-Simons Foundation, Grant \#2019-1403. A portion of this research was carried out at the Jet Propulsion Laboratory, California Institute of Technology, under a contract with the National Aeronautics and Space Administration (80NM0018D0004).
\end{acknowledgments}
\bibliography{references}{}
\bibliographystyle{aasjournal}

\appendix
\section{Robustness of SVD}\label{sec:svd}
\begin{figure*}[t]
    \centering
    \includegraphics[width=0.95\textwidth]{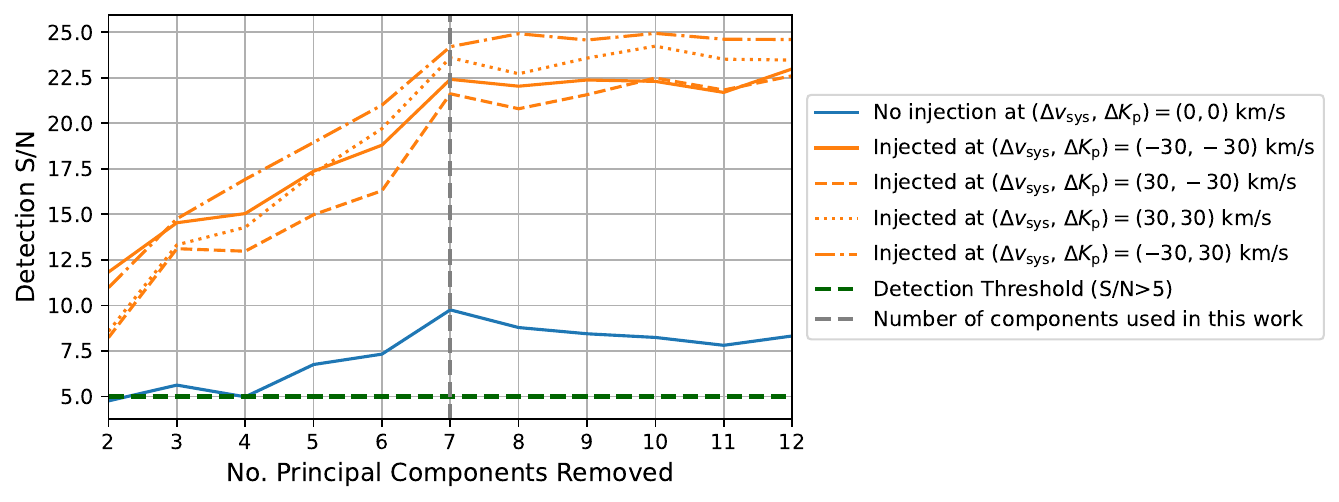}
    \caption{Detection S/N as a function of number of removed principal components. The blue line shows the detection S/N for the observed data without model injection, while the orange lines show the injected data for four different velocity offsets. The dashed green line indicates a detection threshold of $S/N>5$.}
    \label{fig:svd_test}
\end{figure*}

To verify the robustness of our main results against various choices of removed components using SVD, we conducted additional injection-recovery tests. We injected our highest S/N model (fiducial without titanium) into the data at a nominal planet-to-star flux ratio at four velocity offsets ($\Delta$\vsys, $\Delta$\kp). We then calculated the S/N at the injection site, and the results are illustrated in Figure~\ref{fig:svd_test}. The retrieved S/N increases up to seven components and then levels off, which mirrors the behavior of the actual atmospheric signal when no artificial signal is added. This indicates that using seven components effectively cleans the data while mostly retaining the exoplanet's signal. Furthermore, we found that a wide range of components removed still maintains detection in the real data above a conservative S/N threshold of five. Thus, we argue that the main results of this study are robust against the choice of removed principal components.
\section{Additional figures}
The next pages contain additional figures to support various parts of this work.
\begin{figure*}
    \centering
    \includegraphics[width=0.83\textwidth]{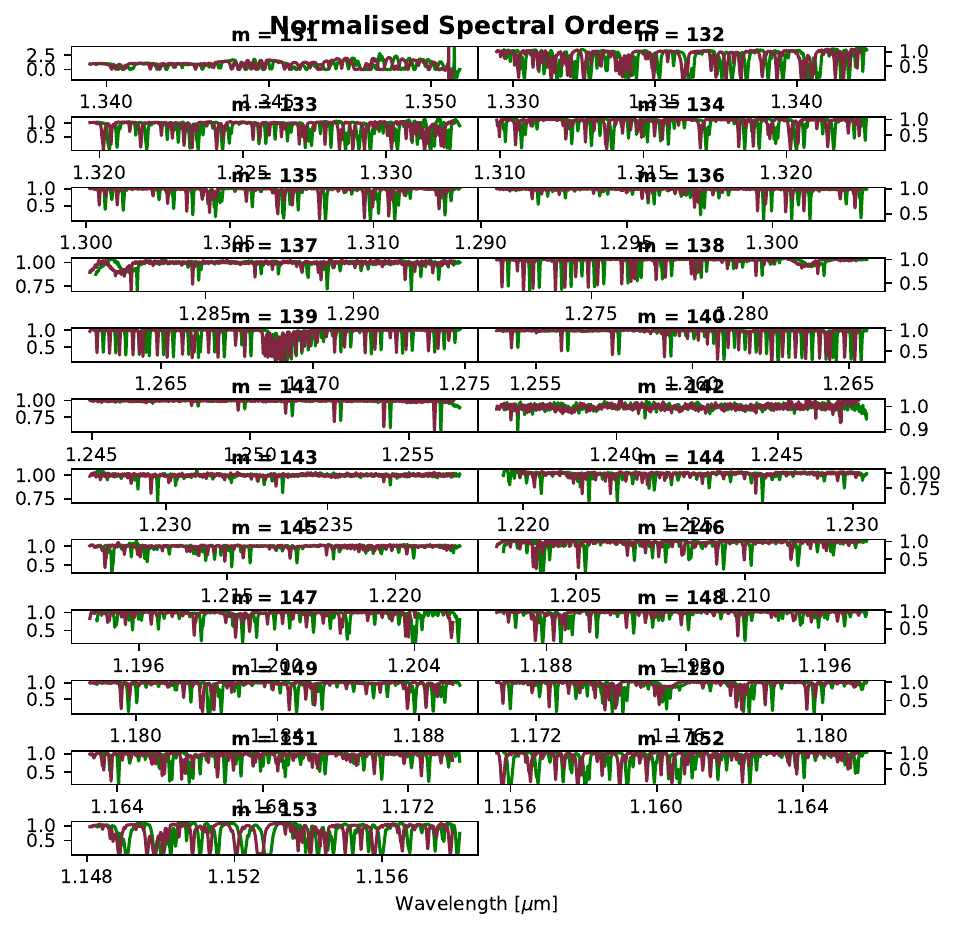}
    \caption{Average obtained normalized spectrum for each spectral order and observing night. The nights have been offset in wavelength for visual purposes. The first night is shown in green and the second night is shown in wine red.} 
    \label{fig:normalised_spectral_orders} 
\end{figure*}
\begin{figure*}
    \centering
    \includegraphics[width=0.34\textwidth]{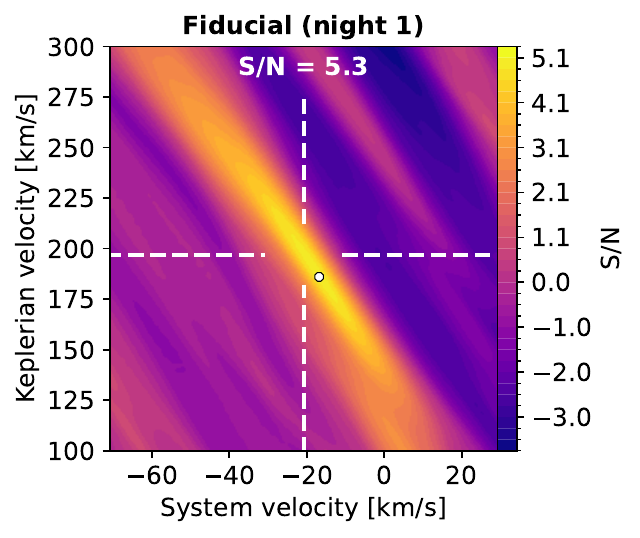}
    \includegraphics[width=0.34\textwidth]{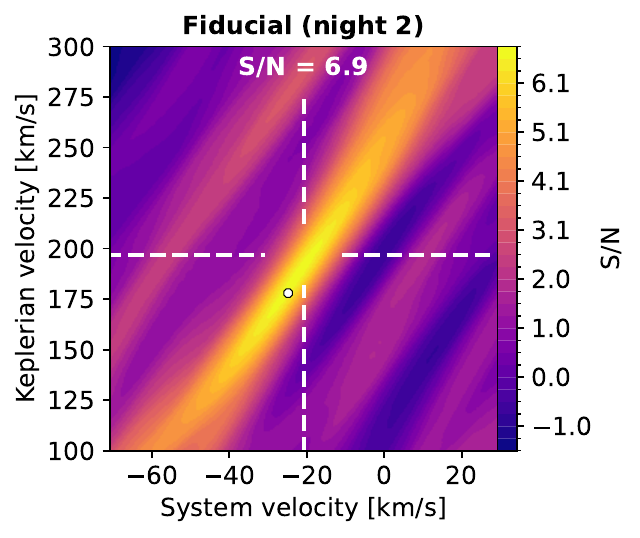}
    \includegraphics[width=0.34\textwidth]{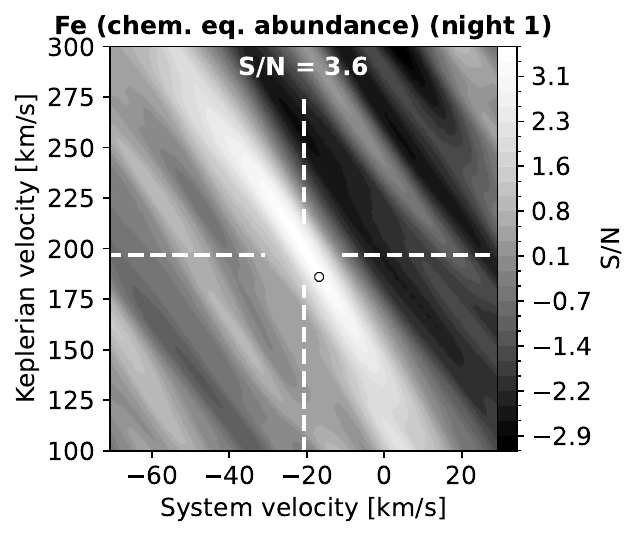}
    \includegraphics[width=0.34\textwidth]{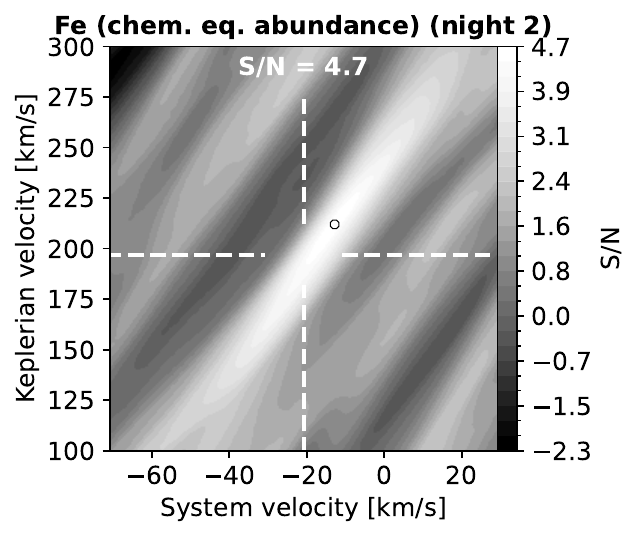}
    \includegraphics[width=0.34\textwidth]{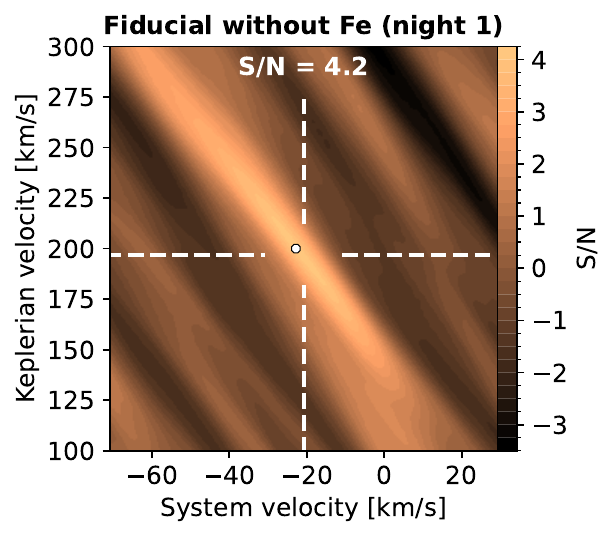}
    \includegraphics[width=0.34\textwidth]{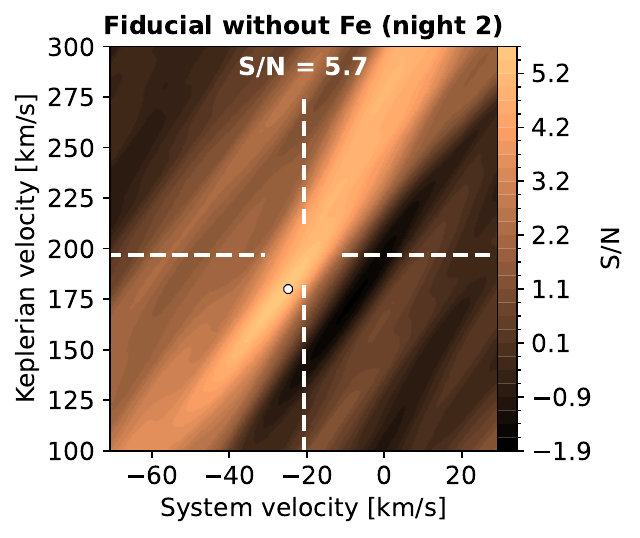}
    \caption{Similar to Fig~\ref{fig:snr_detections}, but instead shown for each night: for the fiducial case, neutral iron (assuming equilibrium chemical abundances), and the fiducial model without iron.} 
\label{fig:snr_detections_each_night}
\end{figure*}
\begin{figure*}
    \centering
    \includegraphics[width=0.89\textwidth]{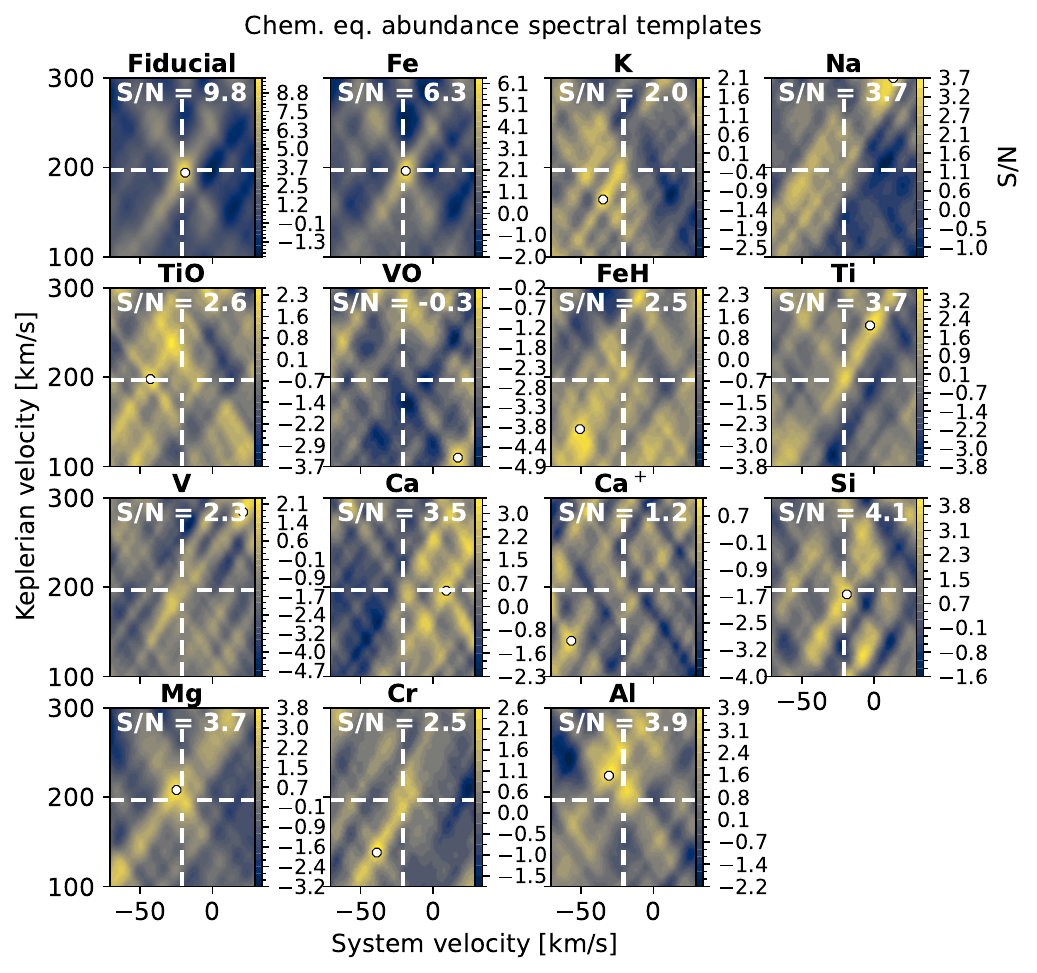}
    \caption{Similar to Fig~\ref{fig:snr_detections}, but shown for all the atmospheric species that have insignificant presence or absence.}
    \label{fig:snr_other_species}
\end{figure*}
\begin{figure*}
    \centering
    \includegraphics[width=0.85\textwidth]{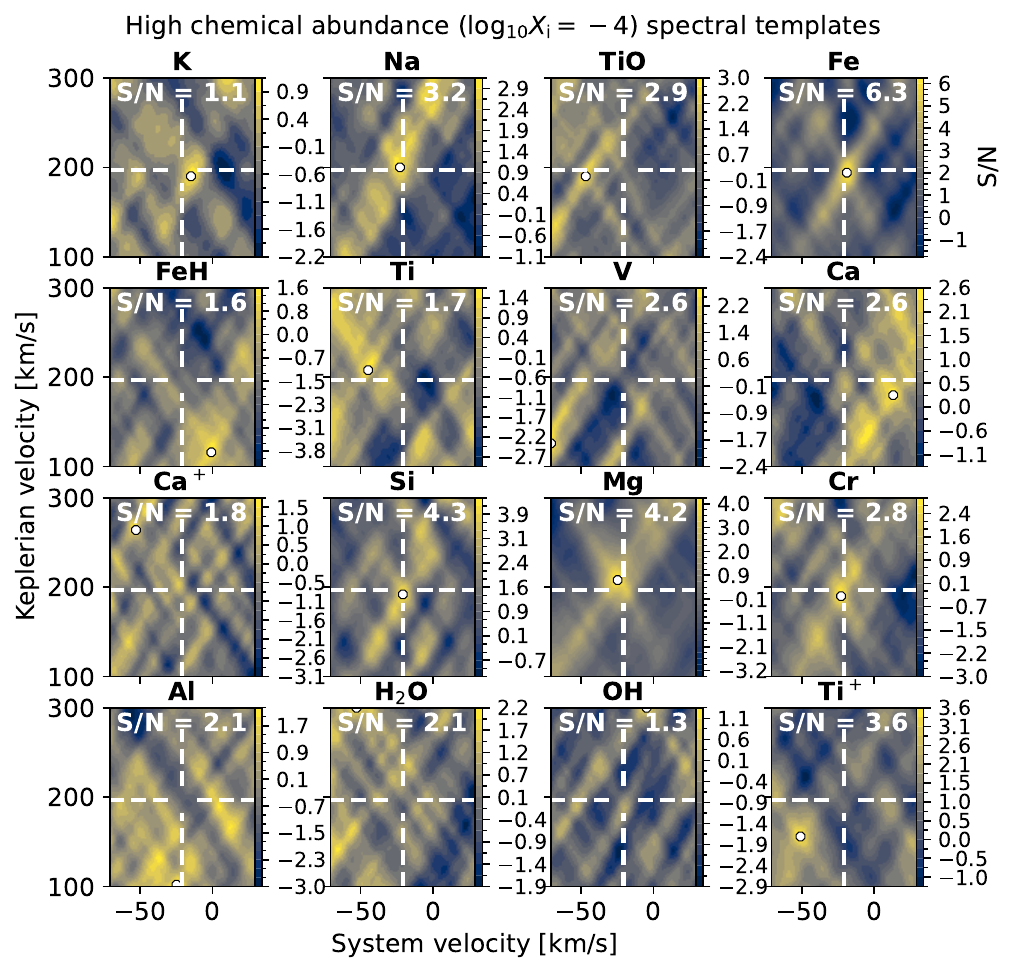}
    \caption{Similar to Fig~\ref{fig:snr_detections}, but shown for all spectral templates that included a single opacity source at an artificially high chemical abundance of $\log_{\rm{10}}VMR=-4$.}
    \label{fig:snr_high_chem_abund}
\end{figure*}
\begin{figure*}
    \centering
    \includegraphics[width=0.85\textwidth]{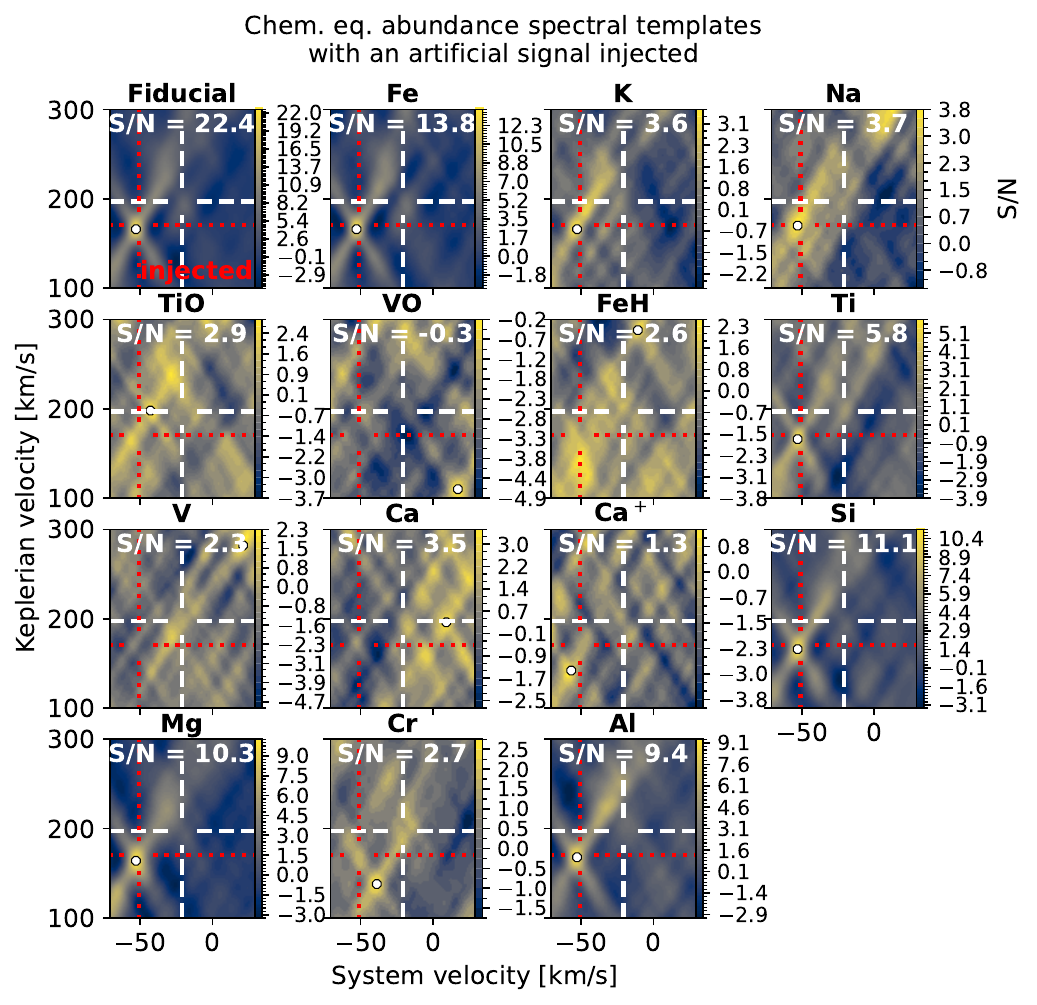}
    \caption{Similar to Fig~\ref{fig:snr_detections}, but shown for all the atmospheric species with an artificial exoplanet signal injected at a velocity offset of $\Delta$\vsys$=\Delta$\kp$=-30$ km /s at nominal scale $a=1$. Recovery of the injected signal depends on the atmospheric species involved, where titanium (Ti) and iron (Fe) have the highest $S/N$ detections at the injected velocity offset.}
    \label{fig:snr_injected}
\end{figure*}
\begin{figure*}
    \centering
    \includegraphics[width=0.85\textwidth]{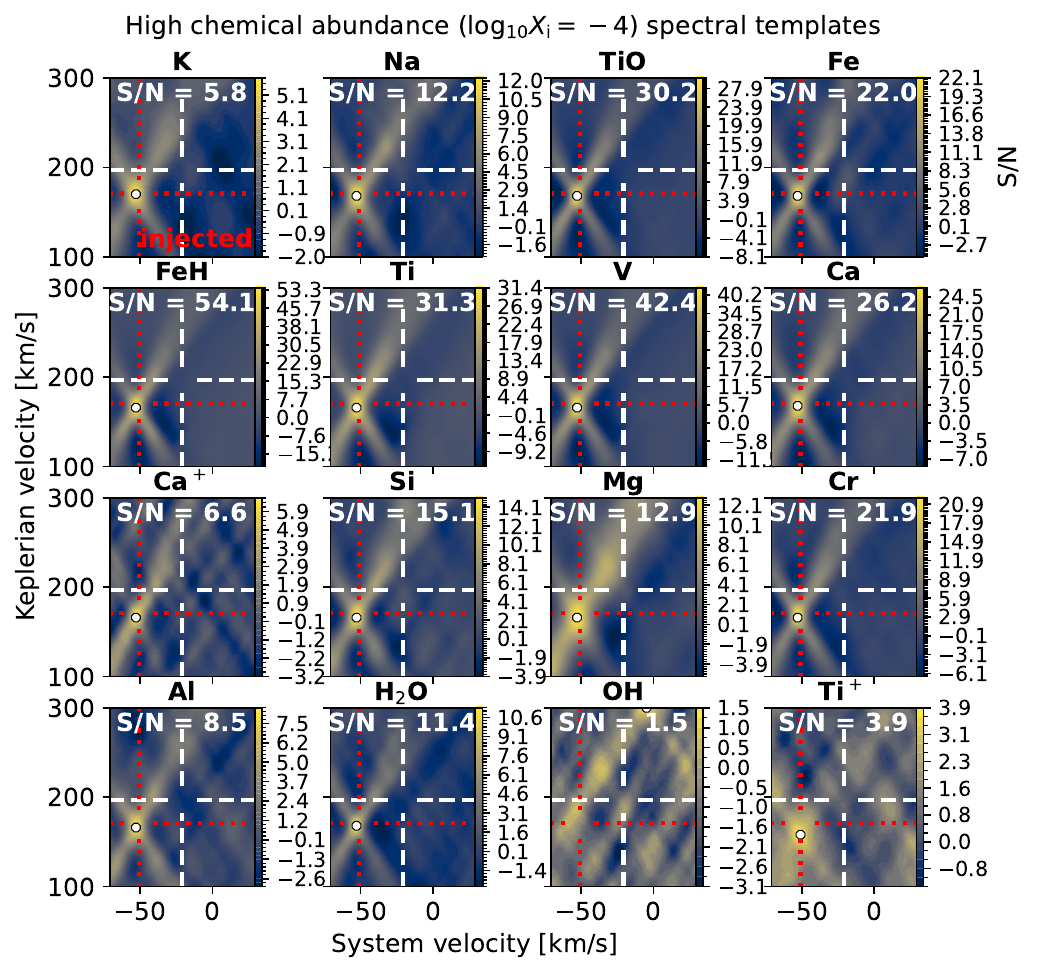}
    \caption{Similar to Fig~\ref{fig:snr_injected}, but shown for all spectral templates that included a single opacity source at an artificially high chemical abundance of $\log_{\rm{10}}VMR=-4$.}
    \label{fig:snr_high_chem_abund_inj}
\end{figure*}
\end{document}